\documentclass[fleqn,usenatbib]{mnras}

\usepackage{newtxtext,newtxmath}

\usepackage[T1]{fontenc}

\DeclareRobustCommand{\VAN}[3]{#2}
\let\VANthebibliography\thebibliography
\def\thebibliography{\DeclareRobustCommand{\VAN}[3]{##3}\VANthebibliography}

\usepackage{graphicx}	
\usepackage{amsmath}	

\usepackage{tikz}
\usetikzlibrary{arrows.meta, positioning, calc}
\tikzset{
    block/.style={
        draw, rectangle, rounded corners,
        minimum height=1cm, minimum width=2.8cm,
        align=center, font=\small
    },
    conv/.style={block, fill=blue!15},
    fuse/.style={block, fill=green!15},
    fc/.style={block, fill=orange!20},
    scalar/.style={block, fill=purple!15},
    concat/.style={
        draw, circle, inner sep=1.5pt, fill=black
    },
    arrow/.style={
        ->, thick
    }
}

\title[SBI multi-temperature ICM]{ERGO-ML and how to directly infer the multi-temperature nature of intra-cluster gas: simulation-based inference with TNG-Cluster}

\author[J. Braspenning et al.]{
Joey Braspenning,$^{1}$\thanks{E-mail: jobraspenning@mpia.de}
Annalisa Pillepich$^{1}$,
Marine Prunier$^{1,2,3}$,
Dimitris Chatzigiannakis$^{1, 4}$,
Dylan Nelson$^{5,6}$
\\
$^{1}$Max-Planck-Institut für Astronomie, Königstuhl 17, D-69117 Heidelberg, Germany \\
$^{2}$D\'epartement de Physique, Universit\'e de Montr\'eal, Succ. Centre – Ville, Montr\'eal, Qu\'ebec, H3C 3J7, Canada\\
$^{3}$Centre de recherche en astrophysique du Quebec (CRAQ)\\
$^{4}$Fakultät für Physik und Astronomie, Universität Heidelberg, Im Neuenheimer Feld 226, D-69120 Heidelberg, Germany\\
$^{5}$Universit\"at Heidelberg, Zentrum f\"ur Astronomie, ITA, Albert-Ueberle-Str. 2, 69120 Heidelberg, Germany\\
$^{6}$Universit\"at Heidelberg, Interdisziplin\"ares Zentrum f\"ur Wissenschaftliches Rechnen, INF 205, 69120 Heidelberg, Germany 
}

\date{Accepted XXX. Received YYY; in original form ZZZ}

\pubyear{\the\year{}}

\begin{document}
\label{firstpage}
\pagerange{\pageref{firstpage}--\pageref{lastpage}}
\maketitle

\begin{abstract}
Though galaxy clusters are virialized and mostly in thermodynamic equilibrium, their internal gas thermodynamics are complex. The hot gas is multiphase and spatially inhomogeneous. The exact state of the gas contains important information about the internal physics governing clusters, however, measuring the multiphase distribution of the X-ray emitting gas is hard if not impossible with current observational methods. In this work (part of the ERGO-ML series), we present a direct inference method of the multiphase temperature distribution of the gas in the intra-cluster medium (ICM), applicable over a large range of spatial and mass scales. It works using different spectral resolutions (XRISM-like and Chandra-like) and realistic photon noise. Our method is an application of simulation-based inference. We train a neural-network on X-ray spectra in the range 0.1-10 keV, generated for $\approx 10^5$ sightlines of massive clusters ($M_{\rm 200c} = 10^{14.2-15.2}~\mathrm{M_\odot}$) from the TNG-Cluster cosmological MHD galaxy simulation. We show that our method recovers the mass-weighted temperature of both individual sightlines and entire clusters with an average prediction error of $< 0.01 ~\mathrm{dex}$, and the total gas mass within $0.04 ~\mathrm{dex}$. On top of that it gives accurate measures of the fraction of sub- and super-virial gas. Our method is capable of extracting the cooler gas phase even if the flux contribution is 3 orders of magnitude below that of the virial gas phase. We then show that we can accurately recover the spatial distribution of multi-phase gas both in relaxed and merging systems. This opens the way to studies of the multiphase nature of the ICM and the violent physics of feedback, accretion and merging.
\end{abstract}




\section{Introduction}
Galaxy groups and clusters are the largest balls of gas in the Universe. For many years the working assumption has been that they are completely filled by virial temperature gas, except perhaps for the regions close to member galaxies. Recently, there has been an increasing focus on how even the hot, X-ray emitting gas has a multi-temperature distribution within the intracluster medium (ICM). The range of temperatures encountered within the X-ray bright gas in galaxy groups and clusters is not unlike the range of temperatures within the multi-phase CGM of galaxies. Low temperature ($10^4~\mathrm{K}$), non-X-ray, gas has, been observed in the ICM of galaxy clusters, for example diffuse MgII absorbers using background quasars \citep[e.g.][]{anand_cool_2022, fresco_searching_2024}. Observational evidence indicates that cold, star-forming gas can co-exist spatially with hot ($>10^7 ~\mathrm{K}$) X-ray emitting gas even within the ICM of galaxy clusters \citep{jachym_alma_2019, waldron_hst_2023}. Because of the mixing of these two gas phases, one can naturally expect clusters to also host cool ($10^{6-7}~\mathrm{K}$) X-ray emitting gas co-spatial with the virial gas phase.

Observationally, it has been clear for some time that such multi-phase gas is ubiquitous in galaxy clusters. For example, \citet{simionescu_metal-rich_2008} found that a model with a continuous range of temperatures between 0.6-3.2 keV ($7\times 10^6 - 4 \times 10^7 ~\mathrm{K}$) gives a better fit to the X-ray spectrum of a cluster core than any model with just one or two temperatures. In cool-core clusters, \citet{olivares_ubiquitous_2019,olivares_h-x-ray_2025} found that cool filaments and star formation are omnipresent, reinforcing the view of multiphase gas over many orders of magnitude in temperature. In fact, the X-ray bright IGM has a spread in temperatures of 4 orders of magnitude between $10^5~\mathrm{K}$ and $10^9~\mathrm{K}$ \citep{bohringer_x-ray_2010}. 

The theoretical realisation that multi-phase distributions can exist across all temperatures, and not just in the ISM, first came in the form of idealised cloud-wind interactions, where it has been found that the cooler gas content can grow over time, although in a fragmentary fashion \citep{gronke_is_2020, gronke_growth_2018, mccourt_characteristic_2018}. Applications to infalling galaxies within galaxy groups and clusters have shown this phenomenon to persist on larger scales \citep{tonnesen_its_2021}, where jelly-fish galaxies can give rise to tails of cold gas clumps, and filamentary cool X-ray emitting gas. Furthermore, the presence of $10^4~\mathrm{K}$ gas in the ICM is reproduced in the TNG-Cluster simulations \citep{staffehl_abundance_2025}.
However, unlike in the CGM, this temperature distribution within the X-ray emitting gas has not yet been widely appreciated and X-ray observations struggle to capture it. This is large due to the virial temperature gas dominating the X-ray emission. This, combined with the, up to recently, low spectral resolution of X-ray instruments, made it near impossible to detect any hot gas with a sub-virial temperatures and establish the distributions of temperatures in the hot ICM. 

On theoretical explanation of the omni-presence of cold gas in the ICM comes from \citet{voit_global_2017}, who employ a precipitation model of cold gas as a natural consequence of the cycle of AGN activity. Such findings are supported by simulations of isolated galaxy clusters, which find the cycles of AGN feedback to be important for the creation and survival of cold gas \citep[see e.g.][]{gaspari_solving_2013, li_cooling_2015}. Though the temperature ranges considered in the definition of multi-phase gas differ between studies, there seems to be consensus that a wide variety of gas phases can co-exist. Due to the mixing of these phases, a continuous temperature distribution would be expected, independent of the exact multi-phase definition used by each study. 

Despite the progress on multi-phase ICM gas and the obvious observational and theoretical clues to its existence and ubiquity, detecting it remains uniquely challenging. Especially for X-ray telescopes, the dominant emitting phase is, except for a few rare cases, the virial temperature gas. Lower temperature gas might exist, but remains hidden below the signal from this dominant component. The most common approach in X-ray spectroscopy is to fit one or two temperature models. The one temperature model will, by definition, only capture the dominant virial gas phase. An alternative is to fit a Gaussian model, which captures the variation in temperature of gas close to the virial temperature, and has been shown to fit offer an improvement over one-temperature models \citep[e.g.][]{mernier_abundance_2015}. None of these methods, however, are able to capture the 2-3 orders of magnitude in temperature between the virial temperature and the lowest temperature which would be X-ray observable (typically $\approx 10^5 ~\mathrm{K}$). 

Cosmological hydrodynamical simulations have only recently reached the cosmological volumes required to have a representative sample of galaxy groups and clusters, but confirm that multi-phase gas is omnipresent, especially in the X-ray regime ($T > 10^5 ~\mathrm{K}$) and poorly characterized by one- or two-temperature models \citep{chatzigiannakis_temperature_2026}. 

Machine learning for analysing spectra has been common place in remote sensing for many years, combing through spectra to classify patches of the observed earth beyond anything humans can do \citep[see review from][]{chen_deep_2014, kattenborn_review_2021}. Beyond that it has become a popular tool in the retrieval of chemical species from spectral observations of exoplanet atmospheres \citep[see e.g.][]{marquez-neila_supervised_2018, cobb_ensemble_2019}, thermodynamics from solar observations \citep{sainz_dalda_recovering_2019}, and stellar classification \citep{sharma_application_2020}. Recently there have also been many applications of neural networks to the spectra of stars, galaxies and quasars \citep[e.g.][]{rastegarnia_deep_2022, rozanski_transformerpayne_2025, zhong_galaxy_2025}  Early on, it was already recognized that such approaches have to be well tested; for example \citet{fadely_star-galaxy_2012} points out how a well trained machine learning model outperforms any template-fitting approach in the classification of stars and galaxies from wide optical surveys, but a poorly trained machine does not.

Within the study of galaxy clusters, machine learning has also started to be a common method. For example, it has been shown that using X-ray images the halo mass can be predicted fairly accurately \citep{ntampaka_dynamical_2018, krippendorf_dynamical_2018, ho_dynamical_2018}. This is usually done in an effort to avoid the problem of mass determination from X-ray density and temperature profiles, which is known to result in biases \citep[e.g.][]{ettori_mass_2013, ghirardini_universal_2019, gianfagna_study_2023, jennings_halo_2023, munoz-echeverria_galaxy_2024, braspenning_hydrostatic_2025}. However, none of these methods directly use the X-ray spectra, nor are they aimed at extracting the ICM properties from X-ray observations.

Modern cosmological hydrodynamical simulations have opened the way to the study of galaxy groups and clusters in their full cosmological context. Simulations such as FLAMINGO \citep{schaye_flamingo_2023, kugel_flamingo_2023} and TNG-Cluster \citep{nelson_introducing_2024} allow for the study of a statistically significant sample of galaxy clusters at different resolutions and with different physics models. They have been shown to reproduce both integrated properties of galaxy groups and clusters, as well as a range of spatially resolved properties such as cool-cores \citep{lehle_heart_2024,braspenning_flamingo_2024}, turbulence \citep{truong_x-ray-inferred_2024}, and X-ray cavities \citep{prunier_x-ray_2025-1, prunier_x-ray_2025}. Such simulations naturally produce multiphase gas \citep{rohr_cooler_2025}, which, however, is sub-dominant both in mass and common observational tracers.

This work falls within the Extracting Reality from Galaxy Observables with Machine Learning (ERGO-ML) series, where the goal is to use novel techniques to extract previously hidden physical information from observational data \citep{eisert_ergo-ml_2023, angeloudi_ergo-ml_2023, eisert_ergo-ml_2024, chadayammuri_ergo-ml_2026, eisert_ergo-ml_2026}. The overarching project makes use of modern cosmological simulations, to train machine learning models that can then be applied to real data.

In this work we focus on using machine learning (ML) to study the signal and extract more information about the sub-dominant temperature components of the ICM. 
We aim to capture the full temperature distribution ($10^6 - 10^9 ~\mathrm{K}$), directly from an X-ray spectrum ($0.1-10~\mathrm{keV}$). We opt to build a neural network that uncovers, non-linearly, information hidden within the spectrum which reveals the presence of sub- or super-viral gas.
The structure of this paper is as follows: Section \ref{sec:methods} presents the simulations and the neural network methodology, the results both on a halo level, and for individual lines-of-sight are presented in Section \ref{sec:results}, where we show that sub-virial gas can be accurately recovered, and also show the effect of telescope characteristic. In Section \ref{sec:model_explanation} we delve into which part of the spectrum are important for our model predictions, and compare the contributions of different line regions. In Section \ref{sec:discussion} we compare with current observational techniques and discuss the limitation of our model. Section \ref{sec:concolusions} presents the conclusions of this work and an outlook to future directions and applicability.

\begin{figure*}
    \centering
    \includegraphics[width=0.85\linewidth]{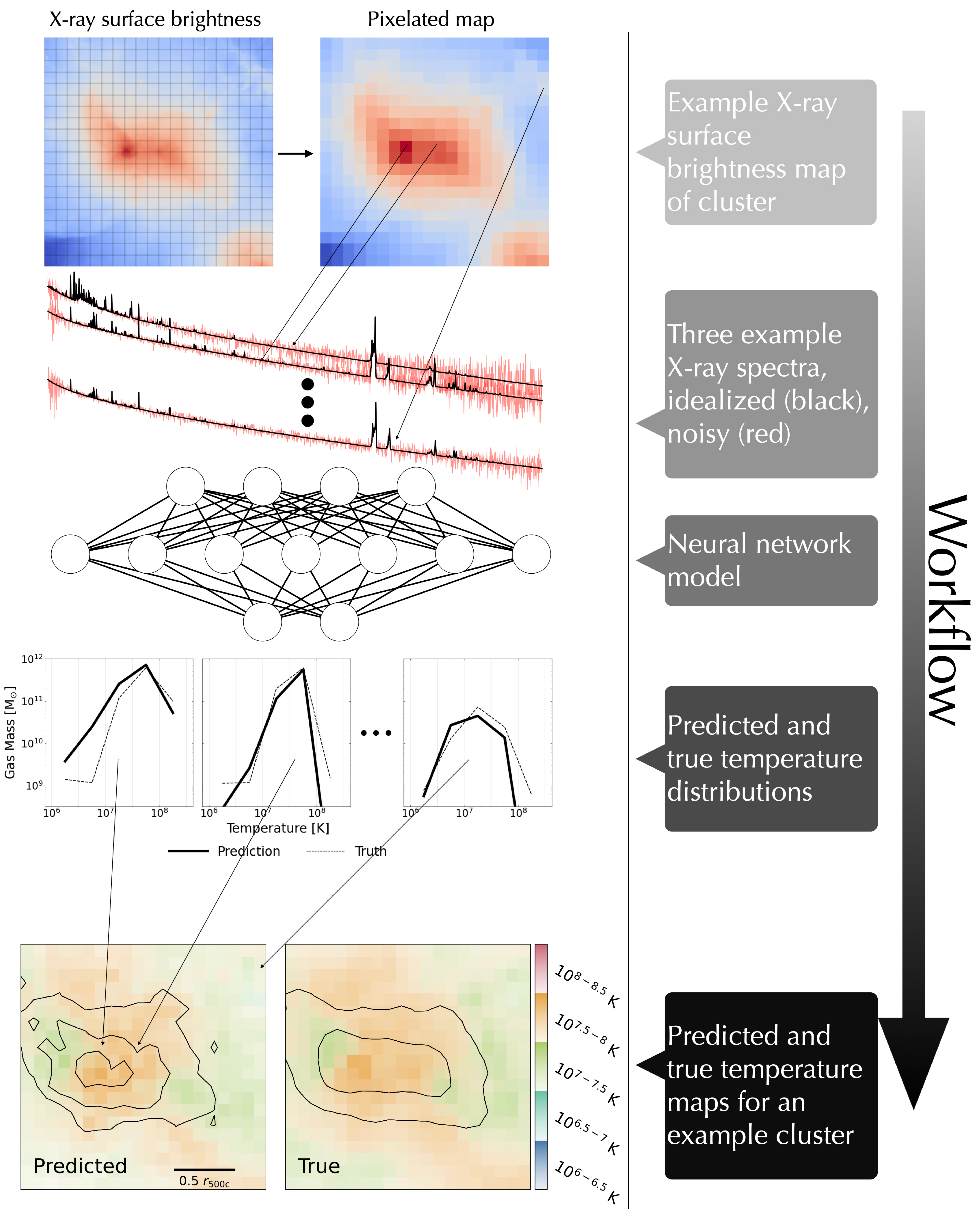}
    \caption{Conceptual flow of the methodology. As a first step, X-ray emission is generated for all gas cells as defined in Section \ref{sec:methods:input_data}, the 2D projection is then divided into 20x20 sight-lines. For each sight-line a summed, total spectrum is generated. We also generate realisitc noise for each spectrum, shown by the red lines in this figure. These spectra are then fed to a neural network, which predicts the temperature distribution in 5 temperature bins for each sight-line. One can take each of these individual predictions, and piece them back together into a predicted multi-temperature map of each halo.}
    \label{fig:concept_figure}
\end{figure*}

\section{Simulation data, reference observations and AI methods} \label{sec:methods}
Conceptually, the methodology exists of 5 steps, these are illustrated in Figure \ref{fig:concept_figure}. First, we take all the gas cells within 2 $R_{\rm 500c}$ of all TNG-Cluster haloes and order them in 400 lines-of-sight. For each line of sight, which is $0.1\times0.1~\mathrm{R_{\rm 500c}}$, we compute for each gas cell the X-ray spectrum. These are projected and added for all gas cells per line of sight. This results in a data cube with a pixelated white-light map, as well as a high-resolution X-ray spectrum in the third direction.
For each sightline we also compute the mass-weighted temperature distribution. Then we train a neural network that takes as input the X-ray spectrum from a sightline, and aims to predict the mass-weighted temperature distribution. Because we have thousands of sightlines across hundreds of clusters, we have a varied training set covering both unimodel, bimodal and flat temperature distributions.
Finally, we apply the trained network to a test set of clusters that have not yet been used for the training. For these clusters we predict the temperature distribution for each sightline, and use those to create 2D maps of the different temperature components for each cluster.

The different steps in this methodology will be described in detail in the following sections.
The cosmological hydrodynamical simulations used in this work are introduced in Section \ref{sec:TNG_cluster}. The data extracted to train our network are described in Section \ref{sec:methods:input_data}. We then describe the X-ray modelling in Section \ref{sec:xray_models}, and the characteristics of the sightlines in Section \ref{sec:sightline_characteristics}. The final two subsections describe the neural network model (Section \ref{sec:architecture}) and the training and hyper-parameter optimisation (Section \ref{sec:training}).

\subsection{TNG-Cluster} \label{sec:TNG_cluster}
TNG-Cluster is a suite of 352 cosmological zoom simulations of massive galaxy clusters ($M_{\rm 200c} > 10^{14.2}~\mathrm{M_\odot}$) selected from a $1~\mathrm{Gpc}$ parent volume \citep{nelson_introducing_2024}. The simulations extend the previous IllustrisTNG simulations \citep{nelson_first_2018, pillepich_first_2018, springel_first_2018, marinacci_first_2018, naiman_first_2018}. In particular, they are designed to complement TNG300, such that the halo mass function of clusters between $M_{\rm 200c} = 10^{14.2}-10^{15.0}~\mathrm{M_\odot}$ is flat, to then follow the mass function of the parent $1~\mathrm{Gpc}$ volume at higher mass. It furthermore has the same resolution ($m_{\rm gas} \approx 10^7~\mathrm{M_\odot}$) as TNG300-1, as well as the same galaxy formation model.

The TNG simulations use the moving-mesh magneto-hydrodynamical code AREPO \citep{springel_e_2010, pakmor_magnetohydrodynamics_2011, pakmor_simulations_2013}. The TNG galaxy formation model includes radiative cooling, star formation, stellar evolution and stellar feedback \citep{pillepich_simulating_2018}, and super massive black hole (SMBH) seeding, merging, and feedback \citep{weinberger_simulating_2017}.

\begin{figure*}
    \centering
    \includegraphics[width=\linewidth]{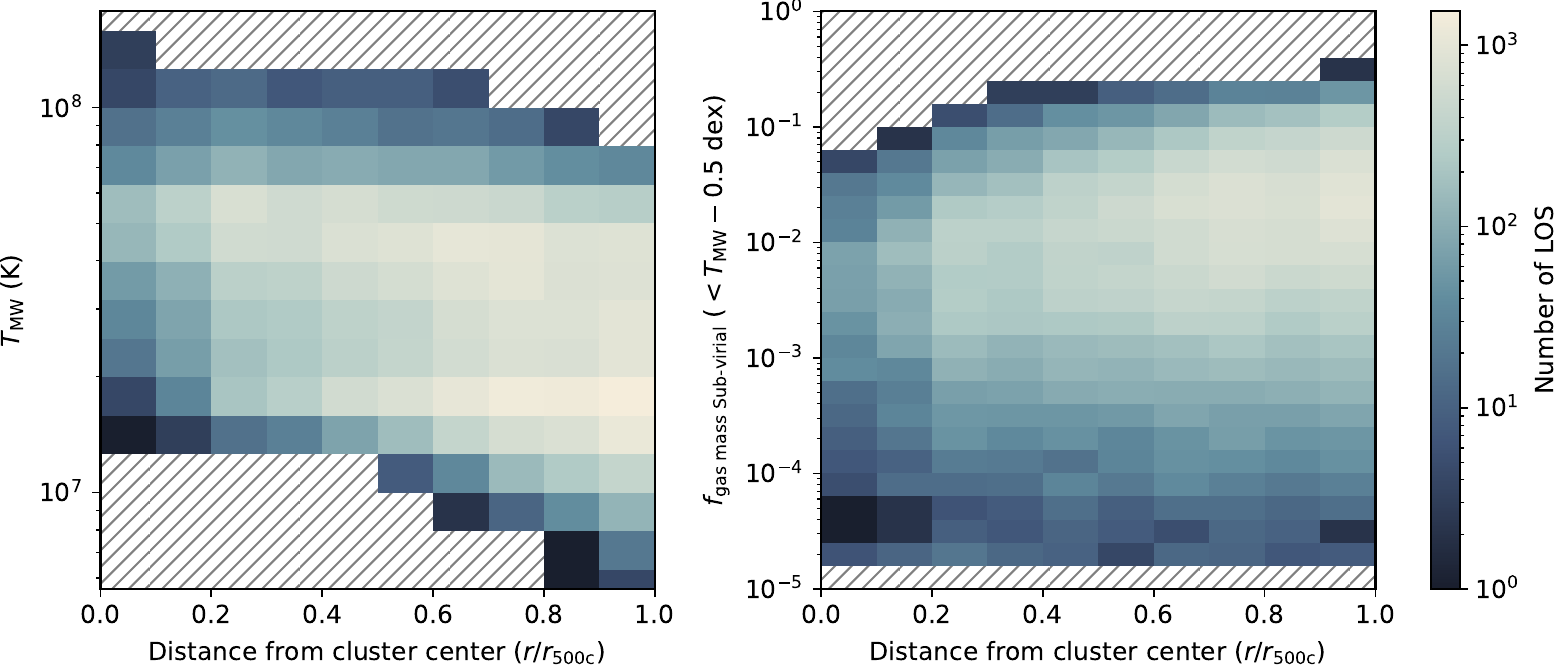}
    \caption{Distribution of line-of-sight summary statistics, showing the spread in properties of our training sample (1/3 of all TNG-Cluster haloes). In each bin the colour indicates the number of sightlines across all input clusters (400 sightlines per cluster). The left panel shows the relation of cluster-centric distance and the mass-weighted temperature of each sightline. As expected, there are more sightlines at large distances, and there is a spread of temperatures from massive groups ($T\sim 10^{7} ~\mathrm{K}$) to clusters ($T\sim 10^{8} ~\mathrm{K}$). The right panel shows the same sightlines and cluster-centric distance, but the second axis is the mass fraction of sub-virial gas in each sightline. We see a clear increase of sub-virial gas towards cluster outskirts, though there is a very large spread. Typical clusters have between 1-10\% sub-virial gas in their sightlines near $R_{\rm 500c}$.}
    \label{fig:sightline_characterstics}
\end{figure*}
\subsection{Sightline-based inference} \label{sec:methods:input_data}
Our aim is to infer the temperature distribution from the X-ray spectra of sightlines of galaxy clusters. To train our machine learning model, we generate X-ray spectra for TNG-Cluster, with the following approach.
For each of the 352 haloes in TNG-Cluster we create 400 lines of sight (sightline). We use the $z=0$ snapshot of each zoom-in cluster and project the gas along the z-axis onto a grid of 20x20 bins of size $0.1~R_{\rm 500c} \times 0.1~R_{\rm 500c}$, corresponding to regions of $\sim (100~\mathrm{kpc})^2$ on the plane of the sky, and creating 400 sightlines within a square of $2\times R_{\rm 500c}$ centered on the cluster core. We define the core as the subfind halo center, which is typically at the center of the brightest/most massive cluster galaxy.

This set-up yields 400 sightlines for each of the 352 clusters, for a total of more than 140,000 sightlines, extracted from the entire extent of the clusters, from the core to the outskirts. Having such a large number of training samples for the neural network is one of the main motivations for splitting the clusters into many sightlines. Another important reason being that this dissection reduces the sensitivity of our method on any specific spatial distribution encoded within the gas from the TNG model, rather, everything gets shuffled.

The mass-weighted temperature distribution is the target distribution to aim to infer with our machine learning method, we measure this distribution for each sightline. We use the mass-weighted temperature because this is directly related to measures of thermodynamic equilibrium and energy balance, unlike emission-weighted temperatures.  Because we are interested in the X-ray emitting gas, we use the mass-weighted temperature between $10^6 - 10^{8.5} ~\mathrm{K}$ in bins of $0.5~\mathrm{dex}$.
For reference, the virial temperature of our clusters sample spans the range of $10^7-10^8 ~\mathrm{K}$, meaning we capture up to $1~\mathrm{dex}$ of sub- and super-virial gas \citep{rohr_cooler_2025, chatzigiannakis_temperature_2026}.

\subsection{X-ray modeling} \label{sec:xray_models}
For each sightline, we create X-ray spectra as described next.
To generate the X-ray spectra, along each sightline we take  all gas cells along a sightline, that satisfy the following three criteria:
\begin{itemize}
    \item[1)] $T_{\rm gas} > 10^{5}~\mathrm{K}$ 
    \item[2)] $\mathrm{SFR} = 0$
    \item[3)] $M_{\rm cell} / M_{\rm target} < 8$
\end{itemize}
the first two criteria ensure that we only select X-ray emitting gas, and the third criterium selects gas that has been refined within the high-resolution region of the TNG-Cluster zoom. Typically, low-resolution gas cells are found outside $2-3~R_{\rm 200c}$ \citep{nelson_introducing_2024}.

We then apply the APEC model from the PyAtomDB package \citep[v3.0.9][]{smith_collisional_2001, foster_updated_2012}. An X-ray spectrum, with a fiducial energy resolution of $5~\mathrm{eV}$ (close to the resolution of X-ray microcalorimeters such as \textit{XRISM} and \textit{NewAthena}) is created between $0.2-10~\mathrm{keV}$ using as input the gas cell temperature, density and metal abundances. We use the 9 abundances as tracked in TNG-Cluster, and set all other abundances to the value corresponding to the metallicity from the "other metals" field in the TNG output. To convert from absolute mass fractions (as tracked by the simulation) to abundances we adopt the solar abundances from \citet{asplund_chemical_2009}. 

Along each sightline we account for the velocity of each gas cell by applying a Doppler shift. As a result, the final spectrum incorporates both thermal (from the gas temperature of individual cells), and kinematic (from the velocities of individual cells) broadening.

\subsubsection{Observational limitations to the spectrum} \label{sec:instrument_response}
In addition to our fiducial setup described so far, we test the limits of our method by degrading the spectrum to mimic observational reality. We consider three different variations:
\begin{itemize}
    \item We apply a fixed spectral bandpass (as opposed to our fiducial $0.2 - 10 ~\mathrm{keV}$ range), excluding energies below $2~\mathrm{keV}$. This can on the one hand be seen as a way of mimicking the close gate-valve on the \textit{XRISM} X-ray telescope, and on the other hand tests the ability of the network to extract information from a limited energy range. This spectral bandpass removes a plethora of spectral lines from cooler gas.
    \item The resolution is degraded. We apply the \textit{Chandra} RMF (Cycle 27) to the input spectrum for each sightline resulting in an average spectral resolution of $120~\mathrm{eV}$ (as opposed to the fiducial $5~\mathrm{eV}$). This is, however, energy dependent and also smooths the spectrum instead of just downsampling, smearing spectral lines, and introducing an effective high-energy cut-off where the \textit{Chandra} RMF becomes insensitive.
    \item We add realistic photon noise for XRISM. To do this, we progressively take the next steps: 1) Convolve with the XRISM response function, 2) Fix the maximum value of the spectrum to result in 500 counts, 3) Sample a Poisson distribution with the expected number of counts for each spectral bin, 4) Deconvolve with the response function.
\end{itemize}

\subsection{Characteristics of the input data} \label{sec:sightline_characteristics}
We quantify the characteristics of the sightlines of the training sample (47,000 out of 140,000 sightlines, belonging to 117 out of 352 clusters in TNG-Cluster) by computing the average temperatures and cluster-centric distances, and by showing the multi-temperature structure. Figure \ref{fig:sightline_characterstics} shows 2D histograms of all the sightlines in the training set , brighter colours indicate more sightlines in a bin. The left panel shows the radial distribution of the sightlines versus the mass-weighted average temperature ($T_{\rm MW}$). By design, there are more sightlines on the cluster outskirts. The average temperatures range between $10^7 ~ \mathrm{K}$ and $10^8 ~ \mathrm{K}$, compatible with our clusters spanning $10^{14.2 - 15.4}~\mathrm{M_\odot}$. A small number of sightlines have super-virial temperatures (for our sample the most massive cluster has a virial temperature of $10^8 ~ \mathrm{K}$), these are distributed evenly across all distances, indicating that both the core and cluster outskirts can be super-virial. The right panel shows the mass fraction of sub-virial gas (at least $0.5~\mathrm{dex}$ below $T_{\rm MW}$) along each sightline. A sightline can have a high average temperature, but still a sizeable fraction of sub-virial gas, making it multi-temperature. Sub-virial temperatures never dominate sightlines. At large distances, the mass fraction of sub-virial gas increases, most likely because of the drop in the radial temperature distribution of clusters \citep[see e.g.][]{lehle_heart_2024, braspenning_flamingo_2024} and sightlines that intersect satellites or cooler clumps in the ICM. At $r > 0.6 ~ R_{\rm 500c}$ a majority of sightlines have between 1-10\% of sub-virial gas, as opposed to $<$1\% at $r < 0.1 ~R_{\rm 500c}$. All of these sightline properties are manifestations of the TNG galaxy formation model, and impacted by modelling choices of stellar of AGN feedback.

\begin{table}
    \centering
    \caption{Hyperparameter choices for the CNN. The learning rate is drawn from $\mathcal{R}$, which indicates the reciprocal (also known as log-uniform) distribution. All other parameters are discrete choices bracketing a range of empirically determined promising values.}
    \begin{tabular}{l|l}
        Hyperparameter name & Values  \\ \hline
        Learning rate & $\mathcal{R}(5\times10^{-5}, 5\times10^{-4})$  \\
        Batch fraction & 0.005/0.01/0.02 \\
        Kernel size 1 & 5/10/15 \\
        Kernel size 2 & 5/10 \\
        Kernel size 3 & 3/5 \\
    \end{tabular}
    \label{tab:hyperparameters}
\end{table}

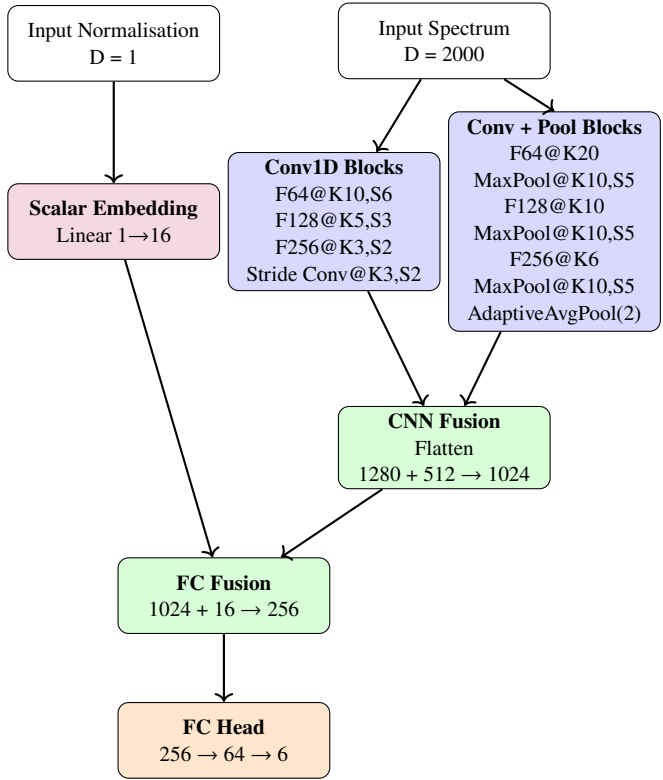
\begin{figure}
\begin{tikzpicture}[node distance=0.9cm and 0.1cm]

\node[conv] (b1) {
\textbf{Conv1D Blocks}\\
F64@K10,S6\\
F128@K5,S3\\
F256@K3,S2\\
Stride Conv@K3,S2
};

\node[conv, right=of b1] (b2) {
\textbf{Conv + Pool Blocks}\\
F64@K20\\
MaxPool@K10,S5\\
F128@K10\\
MaxPool@K10,S5\\
F256@K6\\
MaxPool@K10,S5\\
AdaptiveAvgPool(2)
};

\node[scalar, left =of b1] (scalar) {
\textbf{Scalar Embedding}\\
Linear 1$\rightarrow$16
};

\node[block] (input) at ($ (b1.north)!0.5!(b2.north) + (0,1.2cm) $)
{Input Spectrum\\
D = 2000};
\node[block] (input_norm) at ($ (scalar.north)  + (0,1.85cm) $) 
{Input Normalisation\\
D = 1};

\node[fuse] (cat)  at ($ (b1.south)!0.5!(b2.south) - (0,1.8cm) $){
\textbf{CNN Fusion} \\ 
Flatten \\
1280 + 512 $\rightarrow$ 1024
};


\node[fuse, below left=of cat] (fc1) {
\textbf{FC Fusion}\\
1024 + 16 $\rightarrow$ 256
};


\node[fc, below=of fc1] (fc2) {
\textbf{FC Head}\\
256 $\rightarrow$ 64 $\rightarrow$ 6
};

\draw[arrow] (input) -- (b1);
\draw[arrow] (input) -- (b2.north);

\draw[arrow] (input_norm) -- (scalar);

\draw[arrow] (b1) -- (cat);
\draw[arrow] (b2) -- (cat);

\draw[arrow] (scalar) -- (fc1);

\draw[arrow] (cat) -- (fc1);
\draw[arrow] (fc1) -- (fc2);

\end{tikzpicture}
\caption{Graphical representation of a CNN with fully connected predictive head. This shows one example network architecture with kernel sizes of 10, 5, and 3. The spectrum is treated separately by two convolutional blocks. One focuses on narrow features (i.e. spectral lines): this is the middle block in the figure. The second convolutional block focusses on wide features (i.e. continuum shape): this is the right most block in the figure. The normalisation of the spectrum is embedded with a linear layer. The outputs from the two convolutional blocks are fused and after a single linear layer, there is a further fusion with the embedded scalar, followed by the predictive fully connected head. For different choices of kernel size, the output size of the stride-heavy convolutional block would change, and the input size 1280 of the CNN fusion block could then have a different value.}
\label{fig:nn_architecture}
\end{figure}
\begin{figure*}
    \centering
    \includegraphics[width=\linewidth]{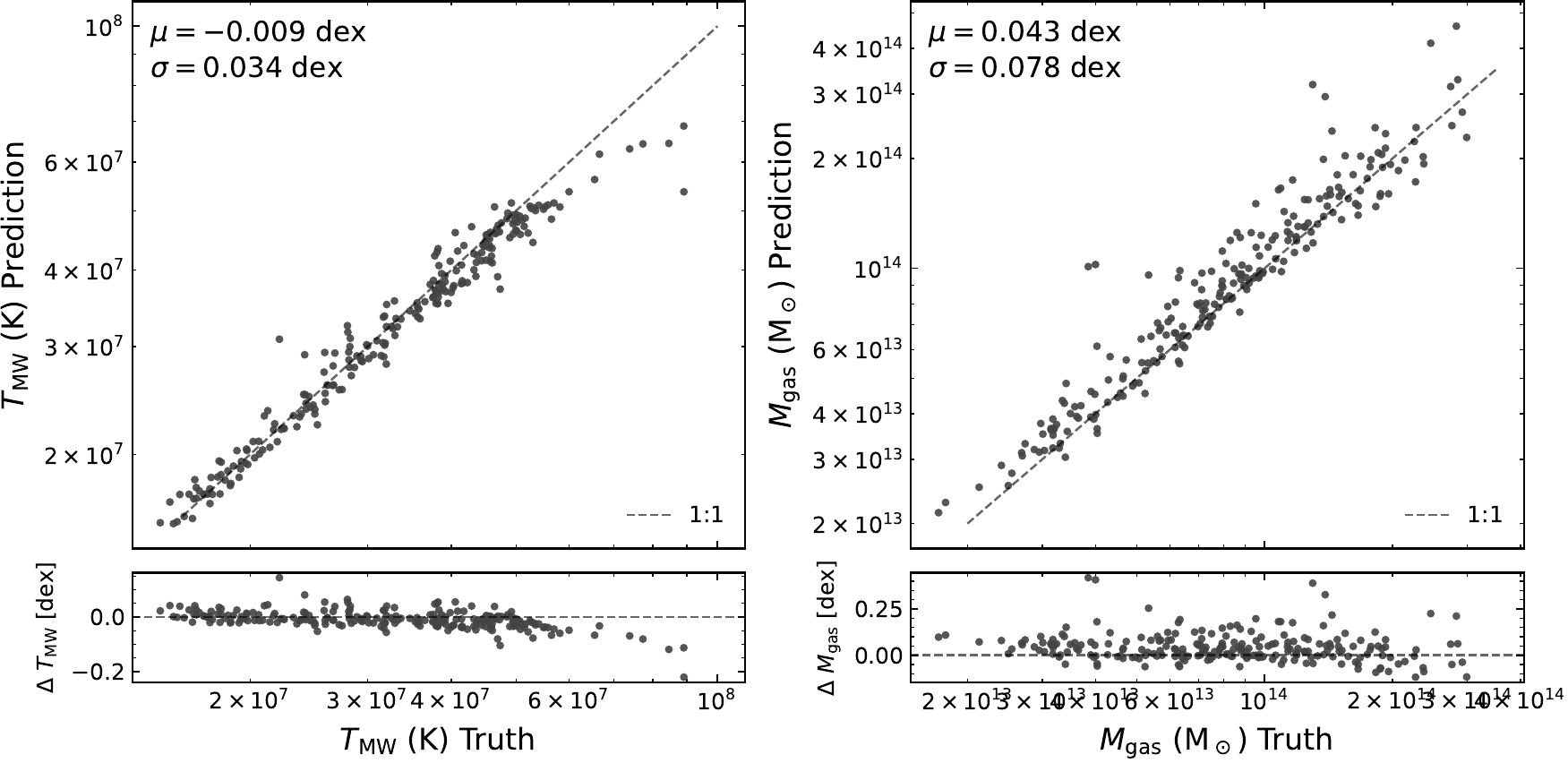}
    \caption{Performance of our method at the cluster-wide level. Here we show the cluster-wide predictions for objects in TNG-Cluster. Predictions for all sidelines belonging to each cluster are collapsed, resulting in a single point. We apply this method to all clusters in our test set (234 for a total of 93,600 sightlines). This results in mass-weighted temperatures (left panels) for each cluster and total gas masses (right panels). In the bottom panels we show the prediction error in dex. The one-to-one line is shown in all panels (dashed line). The temperature predictions are excellent, staying very close to the one-to-one line, showing no bias. The gas mass is slightly over predicted, but shows no trend with halo mass.}
    \label{fig:halo_predictions}
\end{figure*}
\subsection{Neural network architecture} \label{sec:architecture}
The aim of our neural network is to take a raw X-ray input spectrum, and predict the fraction of gas mass in 5 temperature bins, as well as the total gas mass. The input is normalised to have a total integrated flux of unity, the normalisation (i.e. the flux) is passed as a separate input to the network (see below for more details on how it is treated). All the outputs are normalised between 0 and 1, and represent 5 orders of magnitude in gas mass fraction. That is, the minimum fraction that can be predicted for a single temperature bin is $10^{-5}$. The five temperature bins are equally log-spaced between $10^6 - 10^{8.5} ~\mathrm{K}$ ($\approx0.1 - 30 ~\mathrm{keV}$). The predicted total gas mass is similarly normalised between 0 and 1, representing 5 orders in magnitude between $M_{\rm gas} = 10^9 - 10^{14}~\mathrm{M_\odot}$.

The machine learning approach employed in this work is that of a 1D convolutional neural network. We build convolutional layers, with a fully connected regression head. In practise, we settled on two convolutional pathways: one focussing on localized information (such as emission lines), the other optimised to have acces to the global information in the spectrum (such as the shape of the continuum). The input to the network is the X-ray spectrum of a single sightline. We feed the spectrum to the two separate convolutional pathways, before concatenating their outputs, adding information about the normalisation, and feeding it to the regression head, which then leads to a prediction of the temperature distribution and gas mass along the sightline. 

The first pathway, optimised to extract localized information, achieves this by the use of a stride-heavy CNN without pooling. Initially, three convolutional layers, with decreasing kernel sizes, build from 1 to 256 filters, with strides equal to half the kernel size. This results in a rather large output dimensionality, since no pooling is applied. Dimensionality reduction is done due to successive convolutional layers without change in number of filters, but with fixed stride size, until an output size between 2 and 7 is reached. The exact size depends on the initial kernel sizes, which are hyper-parameters of the model.

The second convolutional pathway, optimised for global information, has three convolutional layers, each having kernels twice as large as those used in the first pathway. No strides are used, but after each convolutional layer a MaxPool operation with kernel size 10 and stride length 5 is applied. Finally, after the third convolutional layer, we use an Adaptive Average Pooling operation to reduce the dimensionality to 2.

The result is between 2-7 dimensions from the localized pathway, and 2 dimensions from the global feature pathway, each with 256 filters. We then flatten and concatenate the outputs from the two pathways and feed it to a dense layer with an output shape of 1024. Because of the aggressive striding and pooling operations, the input dimension for this dense layer is comparable to the output dimension, avoiding the common bottle-neck problem of flattening overly large convolutional outputs.

Furthermore, we train a small dense embedding network to transform the normalisation of the spectrum (1 input value) to 16 outputs. Having this small embedding network, instead of a raw concatenation, allows more freedom for the network to normalise the scalar input onto the collective manifold of the embeddings, and not have it disrupt the distributions of gradients of the concatenation layer.

The 16 dimensional output is concatenated with the 1024 dimensional output from the combination of the two convolutional pathways, and fed into a 3 layer, dense MLP, reducing it to a 6 dimensional output vector. These 6 dimensions represent 5 gas mass fractions in the 5 temperature bins and the overall gas mass along the line of sight.

Figure \ref{fig:nn_architecture} shows the architecture when the three kernel sizes are 10, 5, and 3. This results in an output from the stride-heavy convolutional pathway of 5$\times$256. The output from the pool-heavy pathway is always 2$\times$256.

\subsection{Training and hyper-parameter optimisation} \label{sec:training}
To optimise the hyper-parameters of the neural network we make use of the ray-tune framework \citep{liaw_tune_2018}, which implements the Tree-Structured Parzen Estimator (TPE) search algorithm \citep{watanabe_tree-structured_2023} through Optuna \citep{akiba_optuna_2019}. TPE is a Bayesian optimisation method that allows hierarchical choices of parameters, i.e. based on the choice of one parameter different subsets of parameters get unlocked and optimised. Hyperparameter choices are listed in Table \ref{tab:hyperparameters}. Optuna is combined with the ASHA scheduler \citep{li_system_2020} which stops hyperparameter trials if they are not promising, the choice being based on the validation sightlines of the neural network. As the sightlines function, we use the mean Huber sightlines per sample \citep{huber_robust_1964}. Adapting early stopping of trials significantly speeds up the hyperparameter search. The best model is taken from the trial reaching the minimum validation sightlines across the entire training time. The weights of the epoch that reached the minimum validation sightlines are used. This is opposed to using the the final validation sightlines, and final weights, which is sometimes done in the literature.

In this work we choose to employ ASHA on the training interations, with a grace period of 50 and a maximum of 100. That means, every trial runs for at least 50 epochs, and at most 200. We find that this yields very reliable results, as the burn-in phase is always less than 50 epochs, often even less than 30 epochs, and the minimum validation sightlines is typically reached around epoch 80. Training trials with hyperparameter combinations that perform poorly are terminated after 50 epochs, whereas promising trials are allowed to train to 200 epochs, or whenever the ASHA algorithm determines that they are no longer competitive. We use a split of 47885 training samples and 4640 validation samples.

\section{Results} \label{sec:results}
In this section we present the performance and accuracy of our neural network method. 
We start by assessing the inference of the temperature and gas mass predictions for clusters. In Section \ref{sec:halo_temperature_predictions} we derive a single mass-weighted temperatures from our predictions, and compare those to the true mass-weighted temperature from the simulation, finding near perfect agreement.
We then look deeper into the quality of the predictions for individual lines-of-sight in Section \ref{sec:sightline_performance}, where we show, for each temperature bin, how well our inference works. We compare these predictions with a method of using near-neighbours and give quantitative measures of which signal-to-noise has to be achieved before our method fails in Section \ref{sec:nearest_neighbours}. 
We then continue with entire haloes, by presenting inferred temperature maps, recovering merger shock heated gas and plumes of super-virial gas in Section \ref{sec:map_predictions}. We then dissect our clusters into distinct regions and in Section \ref{sec:halo_predictions} we delve into at which energies, which temperature range dominates the spectrum. 
Finally, in Section \ref{sec:sattelitle_predicitons} and Section \ref{sec:cold_gas_profiles} we show that our method is also able to accurately recover the cool gas mass in massive satellites, and the profile of sub-virial gas across the entire radial extend of the halo.

\begin{figure*}
    \centering
    \includegraphics[width=\linewidth]{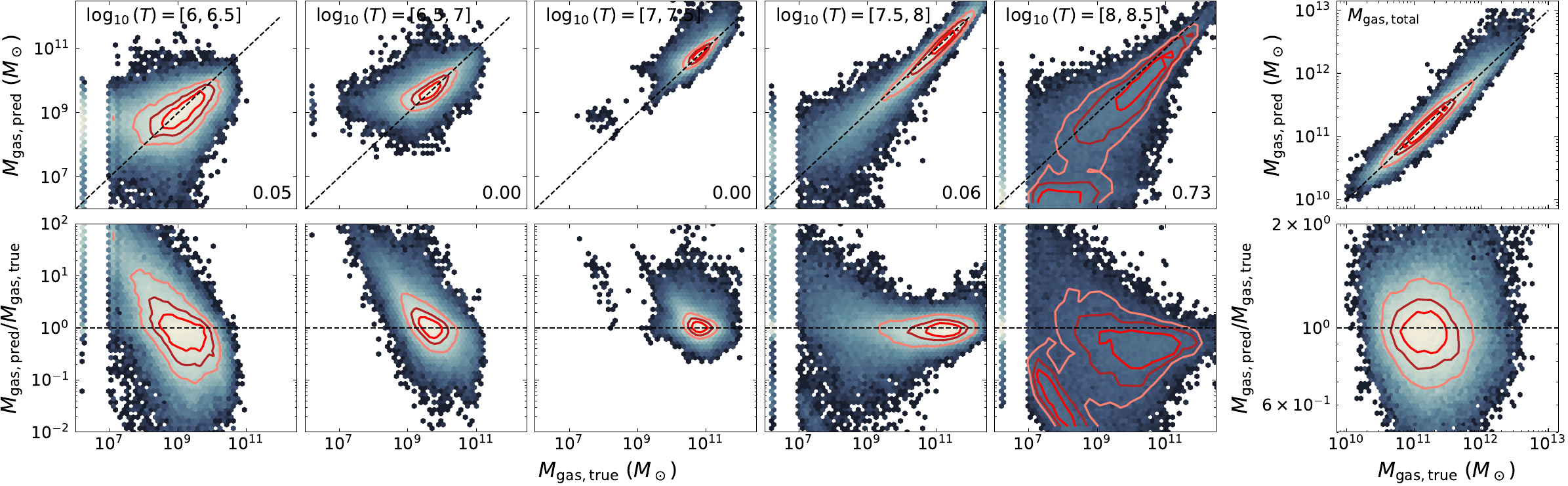}
    \caption{Performance of the neural network for predicting gas masses for each of the five temperature bins, and the total gas mass along all sightlines in TNG-Cluster. All results are for the test set (93,600 sightlines). Left to right the first 5 columns show the performance in the temperature bins, the right-most column show the predicted total gas mass for each sightline. Lighter colours indicate more sightlines, contours show the levels where 25\%, 50\% and 75\% of the sightlines included. The bottom row shows the ratio between the predicted and true gas masses. The one-to-one perfect prediction is indicated by the black dashed line. The numbers in the bottom-right corner indicate the fraction of sightlines with zero true gas mass in each temperature bin. A typical sightline has very good predictions between $10^7$ - $10^8 ~\mathrm{K}$, has a larger scatter, but no overall bias, for temperatures just cooler that that, but has large errors and biases for both the coldest and hottest gas. Looking at the gas mass distributions in the top panels this makes sense, as those temperatures are under-abundant by on average 2-3 orders of magnitude, hence contribute very little to the overall spectrum.}
    \label{fig:error_per_t_bin}
\end{figure*}
\subsection{Cluster-wide temperature and gas mass predictions} \label{sec:halo_temperature_predictions}

We take all the sightlines for each cluster that are within $R_{\rm 500c}$ and combine them to obtain a single mass-weighted temperature and gas mass prediction. Each sightline gives a predicted gas mass in the 5 temperature bins as described in Section \ref{sec:methods:input_data}. Based on the splitting between test and train sets, we use the 234 haloes in the test set, totalling 93,600 sightlines. We compute the mass-weighted temperature based on the predictions in those 5 bins, and compare that with the true mass-weighted temperature of the cluster. The network also predicts the total gas mass along each sightline: we sum all those sightline-based predicted gas masses and compare it with the true gas mass.

Figure \ref{fig:halo_predictions} shows the cluster-wide mass-weighted temperature (left panels) and total gas mass (right panels), for all clusters in the test set. The clusters span temperatures from $10^7 ~\mathrm{K}$ to $10^8 ~ \mathrm{K}$ and total gas masses from $2\times 10^{13} ~ \mathrm{M_\odot}$ to $4 \times 10^{14} ~ \mathrm{M_\odot}$. Temperatures are well recovered over the entire range, proving the efficacy of our methodology. The mass-weighted temperature inference is unbiased and has small errors
\begin{equation}
    \log_{10}T_{\rm MW, pred} - \log_{10}T_{\rm MW, true} = -0.009 \pm 0.034 ~\mathrm{dex}
\end{equation}

The gas mass is slightly over-estimated across the entire mass range, but the recovery is independent of halo mass. 
\begin{equation}
    \log_{10}M_{\rm gas, pred} - \log_{10}M_{\rm gas, true} = 0.043 \pm 0.078 ~ \mathrm{dex}
\end{equation}

These results show that our method is able to recover the mass-weighted temperature from X-ray observations, as well as the gas mass of haloes. Both are recovered with a small error and similarly small spread. 

We note that this is different from other observational approaches which usually struggle to reproduce the true, mass-weighted temperature. Rather, they often try to reproduce an empirical weighting scheme for temperature, the spectroscopic-like temperature, which approximates the biases in temperature estimation from the spectrum. Here, instead of reproducing such a biased temperature, our model extracts the true temperature.

We tried fitting a log-normal to the temperature distribution, but found that it does not describe the temperature in TNG-Cluster well. These findings are in line with \citet{chatzigiannakis_temperature_2026} who find that the log-normal temperature estimate from \textit{XRISM}-like mock observations is similarly biased compared to the mass-weighted temperature.

\begin{figure*}
    \setlength{\lineskip}{20pt}
    \centering
    \includegraphics[width=\linewidth]{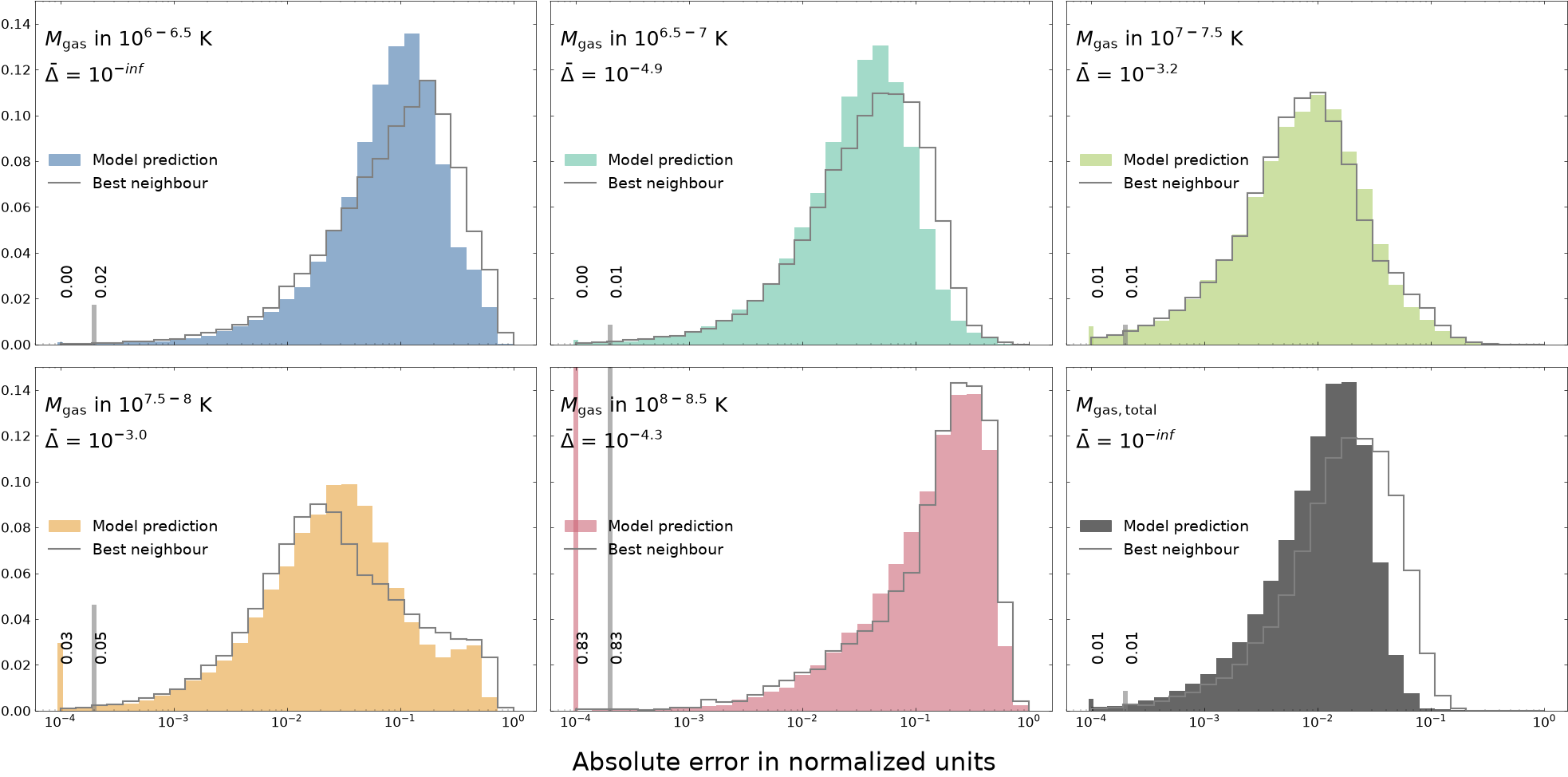} 
    \includegraphics[width=\linewidth]{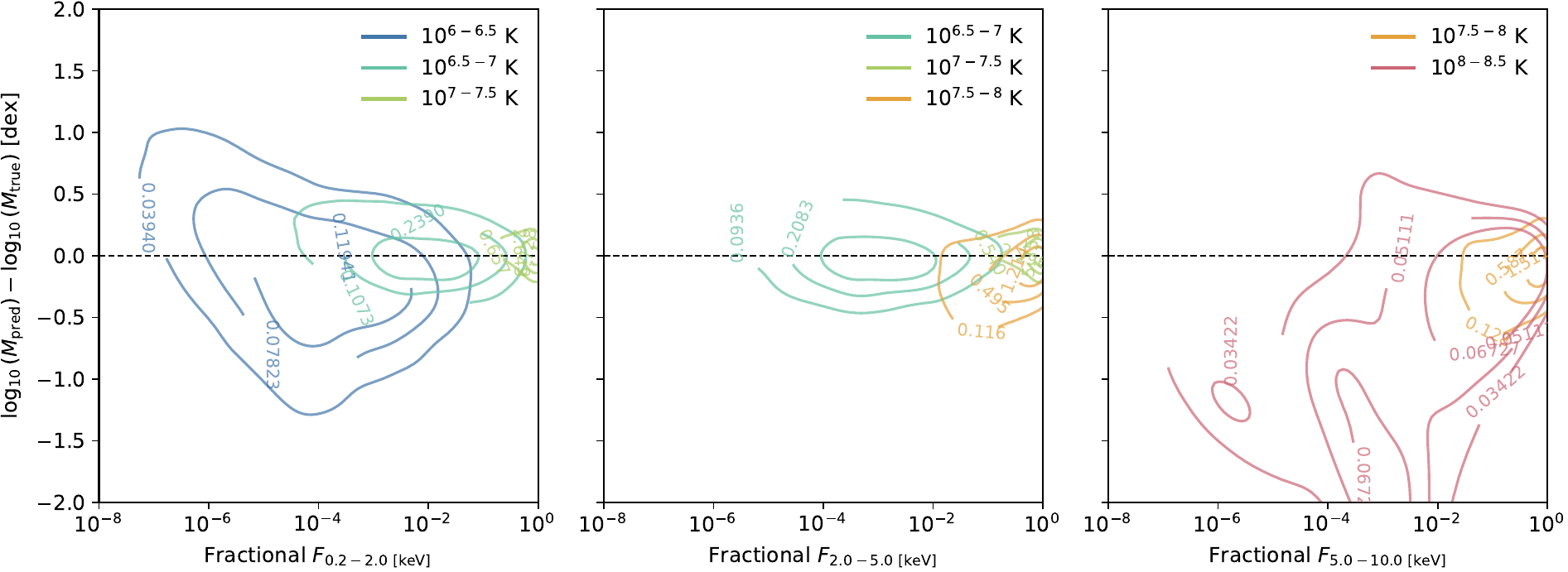}

    \caption{Cross-check of neural network performance against best-neighbours (top) and contribution to the input signal (bottom). The top panels show the five temperature bins (colours) and the gas mass (grey) prediction errors in normalised units. This is the MAE of the neural-network, the black line histogram is the MAE distribution of neighbours with a fixed distance, where the distance is chosen such that the KS-test statistic between the two distributions is minimized. Sightlines with zero prediction error, or zero difference between neighbours are shown as bars on the left-hand side, with the numbers indicating the fraction. These sightlines almost always have no gas in that particular temperature bin, rather than perfectly predicting the non-zero amount of gas. This allows to measure what the average difference in the spectrum is our model can resolve. The value $\bar{\Delta}$ gives the average fractional difference between the two neighbour spectra. We see that even for the worst performing temperature bin, our network can disentangle the underlying temperature distribution of spectra that differ on average by one part in a thousand. 
    The bottom panels show the log absolute error (in dex) for each temperature bin, as a function of the fraction that the gas at that temperature contributes to the total spectrum. From left to right, the panels show the low-, mid-, and high-energy parts of the spectrum, where different temperatures are expected to dominate. Only non-zero predictions are shown, this mostly affects the highest temperature bins, where only 17\% of all sightlines have non-zero values (see top figure).}
    \label{fig:mse_neighbours_network}
\end{figure*}

\subsection{Performance across sightlines} \label{sec:sightline_performance}
Having ascertained that we are able to recover cluster-level temperatures and gas masses, we now look deeper into the inferences for individual sightlines. Since we do not predict a single temperature, but a distribution, we quantify the performance separately for each temperature bin. 
We do this for all the sightlines in our test set. Figure \ref{fig:error_per_t_bin} shows, in the top row the true and predicted gas mass in each temperature bin, as well as the total predicted gas mass across all bins (rightmost panels). The red contours contain 25\%, 50\% and 75\% of the population. Note that what we actually predict is a mass fraction in each temperature bin, which we then multiply by the predicted gas mass. This is compared with the true mass fraction multiplied by the true gas mass. Hence, an error in the total gas mass prediction would propagate to each temperature bin in the same way. This prediction of the total gas mass for each sightline is shown in the final column. From here we see that the total gas mass is always within a factor 2, with the 50\% percentile having a fractional error $< 0.2~\mathrm{dex}$. Looking at the other columns, it becomes clear that the error per temperature bin is not dominated by the gas mass prediction, but rather by the prediction of the mass fraction in each bin. As the gas mass per temperature bin is a much more directly applicable and interpretable quantity, we choose to show that instead of the mass fraction.

For all temperature bins, the majority of sightlines (contained by the red contours) are close to the one-to-one line. The temperature bins where the majority of the gas mass is present, near the virial temperatures of our cluster sample (bin 2: $10^{7-7.5}~\mathrm{K}$, and bin 3: $10^{7.5-8}~\mathrm{K}$) show excellent predictions with the distribution having a narrow shape along the one-to-one line. 
For the highest, super-virial, temperature bin, most sightlines have no gas at this temperature (73\%), demonstrated by the darker colours and wide contours. All sightlines with zero gas mass in this highest bin are placed on the left hand side of the panel. The colour shows that a large fraction of those also has a very low predicted mass. However, for the subset of sightlines with a non-negligible amount of this hottest gas the predictions show a large scatter. 
For the lowest temperature bins ($10^{6-6.5} + 10^{6.5-7}~\mathrm{K}$), we also see a number of sightlines with none of the coolest gas (5\%), but non-zero predictions. In general, the 50\% contours are tight and follow the 1-to-1 relation, however the 75\% contours, and even more so the extended scatter, show a typical regression to the mean. Low gas masses are over-predicted, and high gas masses under-predicted. This is likely due to the very small contribution of this cold gas to the total X-ray spectrum, something that we will explore in more detail in Section \ref{sec:halo_predictions}.

\subsection{Cross-checking performance against nearest neighbours} \label{sec:nearest_neighbours}
Since neural networks are, in practice, nothing more than non-linear interpolation machines, we check how well the prediction from our network compares to using the target values from the Nth nearest-neighbour spectrum. In the latter case, instead of using a neural network, we use another spectrum in the test set at fixed Euclidean distance. We define the distance for each spectrum $i$ in the training set to all other spectra $j$ in the training set as
\begin{equation}
    d^{\rm input}_{i,j} = \sqrt{\sum_{k=1}^{2000}(S_{i, k} - S_{j, k})^2} \, ,
\end{equation}
where $S$ is the normalised logarithmic spectrum, and $k$ is the index in the 2000 dimensional spectrum.

We also compute the distance between the pairs of targets
\begin{equation}
    d^{\rm target}_{i,j,k} = \sqrt{(T_{i, k} - T_{j, k})^2} \, ,
\end{equation}
with $T$ the logarithmic target of 5 temperature bins and a gas mass, indexed by $k$. That is, for two spectra $i$ and $j$ at distance $d^{\rm input}_{i, j}$, the difference in their values for the first temperature bin corresponds to $k=0$, i.e. $d^{\rm target}_{i,j,0}$, and the difference between their values for the total gas mass corresponds to $k = 5$, i.e. $d^{\rm target}_{i,j,5}$.

Next, we define discrete steps in input distance of 0.1 between 0 and 1.0, and find for each spectrum $i$ in the dataset a second spectrum $j$ that is closest to being separated by exactly that distance
\begin{equation}
    j = argmin\left[ \left( d^{\rm input}_{i,j} - d_{\rm select} \right)^2 \right]|_i \, ,
\end{equation}
where $d_{\rm select}$ is our discrete distance steps. Setting $d_{\rm select} = 0$ is equivalent to using the nearest-neighbour spectrum.

Then, for each bin $k$ in the target distribution, we create a histogram of $d^{\rm target}_{i,j,k}$ at fixed distance $d_{\rm select}$. We also create a histogram of the mean-squared error of the network in each target bin $k$ applied to the same spectral dataset. 
We then perform a Kolmogorov-Smirnov (KS) test on this distribution of $d^{\rm target}$ and the distribution of actual mean-squared errors from the neural network. The distance at which the KS test statistic is minimized tells us the level of difference in the spectrum that the network can differentiate between. Crucially, this is done independently for each of the six predicted values. Because there are a large number of sightlines with no gas in some of the temperature bins (especially for the highest temperature bin), and both the network and neighbour difference are excellent at predicting such zeros, we exclude them when doing the KS test. That means, the KS test is only done on the distribution of non-zero errors.

The top 6 panels of Figure \ref{fig:mse_neighbours_network} show the result of this cross-check. Each panel shows one temperature bin, and the final (second row, right) panel shows the gas mass. In each temperature bin, both approaches will have some sightlines where the error is exactly zero, these zero errors are shown as vertical bars on the left-hand side, with the number indicating the fraction of the total sightlines. The highest-temperature bin has no gas for 83 per cent of sightlines, both approaches reproduce this well, resulting in a large number of zero errors for this bin. 
The main metric to measure how our network performs is the average difference between the neighbouring spectra distance for which the KS-test statistic is minimized. This number is denoted by $\bar{\Delta} = d_{i, j}^{\rm input} / N_{\rm spec}$ in the top-left corner of each panel, where $N_{\rm spec} = 2000$ is the number of spectral bins. 
We see that for the lowest temperature bin, as well as for the gas mass prediction, the network performs better than using the nearest neighbour (i.e. $\bar{\Delta} < d_{\text{nearest neighbour}}$, which gives $\bar{\Delta} = 0 = 10^{-\text{inf}}$). The worst performance is found for the fourth temperature bin ($10^{7.5-8.0}~\mathrm{K}$). However, even for that bin, the difference in the spectrum that can be differentiated is still $10^{-3.0}$ per spectral bin. We can rephrase this: Using similar spectra instead of the neural network only becomes better when the signal-to-noise of the matched spectra is better than $\approx 1000$.

To see what level of contribution to the X-ray flux is needed for the network to accurately recover the gas mass in each temperature bin, we now show the relationship between fractional flux in different energy bands and prediction error.
The bottom 3 panels of Figure \ref{fig:mse_neighbours_network} show the fractional flux contributed by each temperature bin on the horizontal axis. The fractional flux is computed in three different energy ranges, $0.2 - 2.0~\mathrm{keV}$ where we expect lower temperature gas to have strong emission lines, $2.0 - 5.0~\mathrm{keV}$ where the virial temperature gas will likely dominate, but a few strong emission lines are still present, and $5.0 - 10.0~\mathrm{keV}$ where there are almost no emission lines and the spectrum becomes dominates by the continuum. For each of these ranges, we show the contours of the prediction errors on the gas mass in each temperature bin. We immediately see that if a temperature bin contributes more than 10\% of the flux, the prediction is unbiased and accurate. However, for the low temperature gas this extends to much lower flux contributions. Even temperature bins that contribute only $10^{-5}$ of the total flux are accurate predicted for gas of $10^{6.5 - 7}~\mathrm{K}$. For the lowest temperature gas, we see and upturn in the prediction error, but only at very low flux contributions.

An exception is the highest temperature bin, but we note that only 17\% of the sightlines have this hot gas present. When the mass of this gas phase is zero, we accurately predict the absence of this gas. However, when it is present, this gas does not have any emission lines, and only comes to dominate the spectrum at the very highest energies. We postulate that the lack of emission lines, and the limited energy range where it has a significant contribution, are the reason for the relatively large errors we see here.

\begin{figure*}
    \centering
    \includegraphics[width=.7\linewidth]{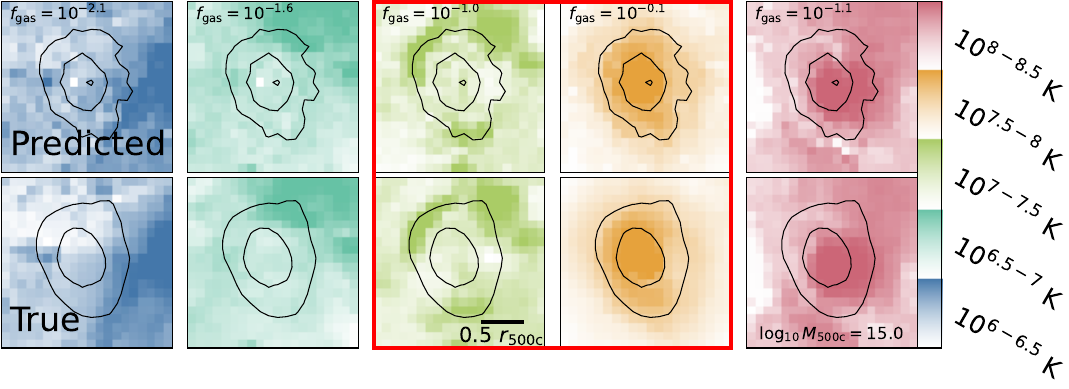}
    \includegraphics[width=.7\linewidth]{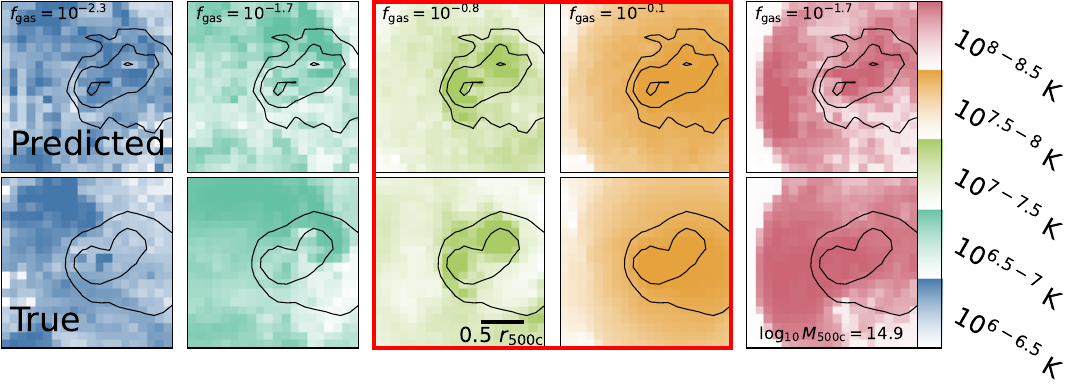}
    \includegraphics[width=.7\linewidth]{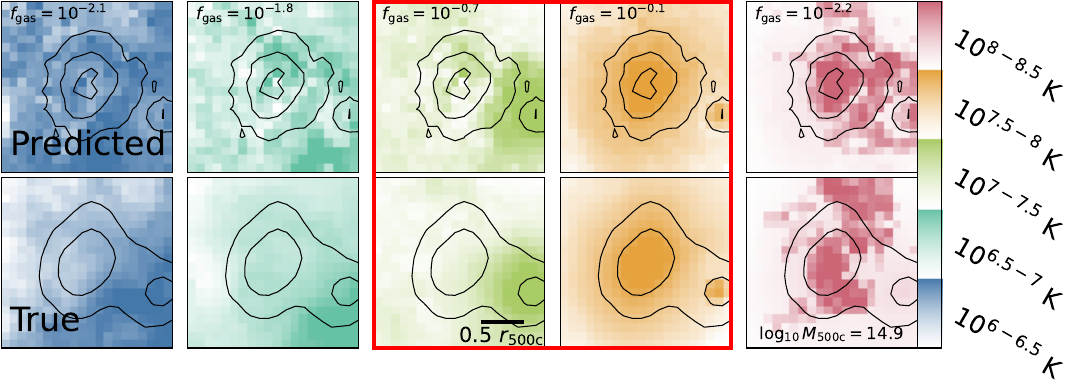}
    \includegraphics[width=.7\linewidth]{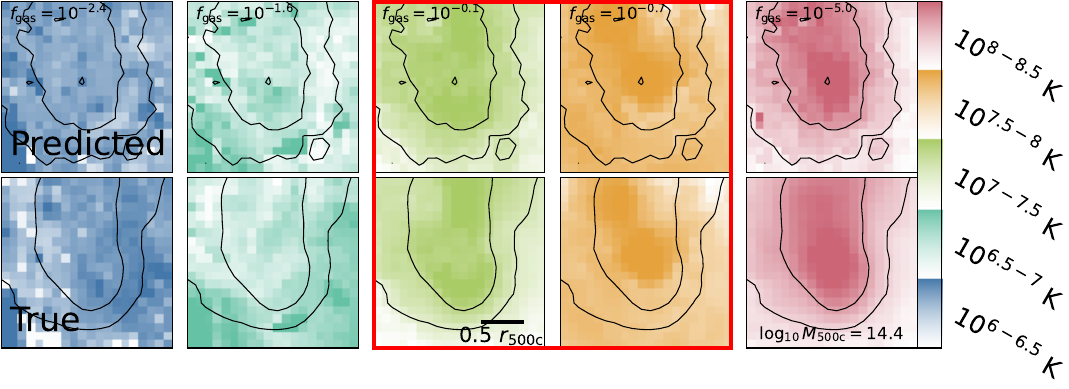}
    \includegraphics[width=.7\linewidth]{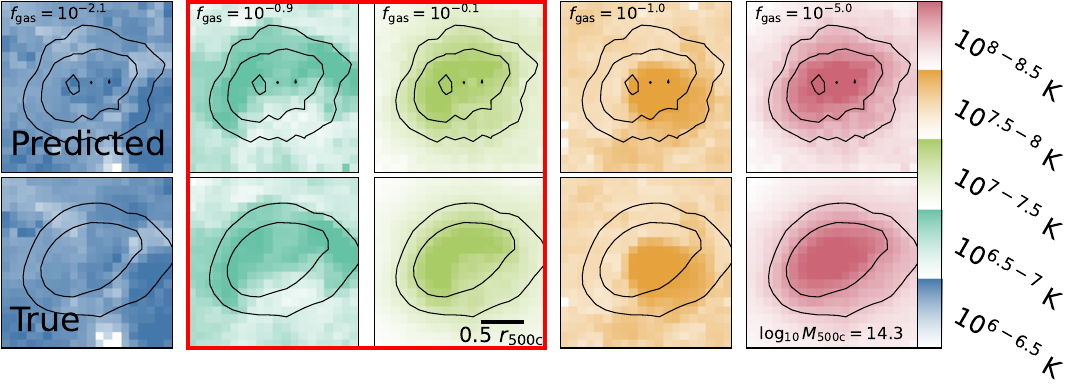}

    \caption{Performance of our method at the map-wide, cluster level. Here we show gas temperature maps for 5 example clusters from the test set showing a diverse range of morphologies. The top rows show the predicted gas mass maps in the 5 temperature bins, the bottom rows show the ground truth. The intensity of the colour encodes the relative abundance of a temperature along each sightline. The red boxes enclose temperatures that harbour at least 10 per cent of the total gas mass of the cluster. The gas fraction in each temperature bin is listed in the top left corners, and the halo mass in the bottom right corner of each set of panels for the clusters. Both more relaxed objects, and merger morphologies are well reproduced.}
    \label{fig:predicted_true_maps}
\end{figure*}
\subsection{Map-level predictions} \label{sec:map_predictions}
Our neural network only sees individual sightlines, without any notion of their location within the 2D projection that we created, or which halo a particular sightline belongs to. As such, all correlation between neighbouring pixels is completely broken. An interesting test, and important application, is whether the predicted total gas mass and gas mass per temperature bin show any reasonable pattern when putting all sightlines of the same cluster together. This complements the results shown in Section \ref{sec:halo_temperature_predictions}, where we have shown that the mass-weighted temperatures of clusters and the total gas mass are well reproduced.

In Figure \ref{fig:predicted_true_maps} we show the temperature maps for 5 clusters from the test set. For each cluster we show the predicted gas mass in each of the 5 temperature bins in the top row, and the true value in the bottom row. For each temperature bin, the colour shows the amount of gas in each sightline normalised to the sightline of that cluster with the maximum gas mass in that temperature bin. At the top of each column, we indicate the fraction of the total gas mass of the cluster that falls into that temperature bin. The panels corresponding to bins containing at least 10 per cent of the total gas mass of the cluster are outlined in red.
The black contours show the gas mass surface density, either true or predicted.
At first glance, the predicted and true maps show excellent agreement. 

Turning our attention to the different haloes, the first three are massive objects ($\sim 10^{15}~\mathrm{M_\odot}$), whereas the latter two have lower masses ($10^{14.3}~\mathrm{M_\odot}$). This mass difference is directly reflected in the dominant temperature component, with the higher mass objects being dominated by higher temperature gas. However, none of these massive haloes have a large fraction of their gas in the highest temperature bin, highlighting once again the challenge in capturing that phase.

More than reproducing just the average temperature, our model also reproduces spatially coherent temperature variations in each temperature bin.
For the first cluster we see a cooler gas phase, containing in total 10 per cent of the total gas mass, wrapping around the hotter core towards the north. Both the intensity and structure of this cooler phase are well reproduced. The even cooler phase, $1~\mathrm{dex}$ below the dominant temperature, and containing just 2.5 per cent of the gas mass, also shows spatially matching predictions, with a clear concentration of this even cooler gas in the north-west.

In the second cluster, we capture a hot merger shock propagating to the west, even though this only contributes 2 per cent of the total gas mass. Furthermore, we capture cooler cores of both the merging haloes in the middle panels.

For the third cluster, which is an early stage merger, we see a clear separation of dominant temperature between the two cores. The central core is dominated by higher temperature gas, with the less massive core to the south-east contains lower temperature gas. The spatial distribution of both these phases is well reproduced.

The fourth cluster, also a merger but at lower total halo mass, shows an elongated structure in the highest 3 temperature bin ($T > 10^{7}~\mathrm{K}$. We specifically not that even though the hottest gas phase only contains a fraction of $10^{-5}$ of the total gas mass, its general morphology is still reproduced.

Finally, the last halo shows a complex structure in its lower temperature phases. In the $10^{6.5-7.0}~\mathrm{K}$ bin we see a cooler phase wrapping around the core. In the next temperature bin $10^{7.0-7.5}~\mathrm{K}$ we see this same front but closer to the core, suggesting a coherent structure over large scales with an inside out temperature gradient. The $10^{7.0-7.5}~\mathrm{K}$ temperature bin traces the core gas, and shows clearly that in the direction of the front, less of this hotter gas is present. We note that all these features are excellently reproduced, making it nearly impossible to distinguish the true and predicted maps.

Our model is clearly capabale of generating not only cluster-wide mass-weighted temperatures, but also the 2D spatial distribution of the 5 different temperature phases we consider. From visual inspection, this generalises well to all clusters. In fact, the merging systems shown in Fig. \ref{fig:predicted_true_maps} represent some of the most disturbed maps in our sample.

\subsection{Contribution of different temperature components to the cluster level predictions} \label{sec:halo_predictions}
\begin{figure*}
    \centering
    \includegraphics[width = \linewidth]{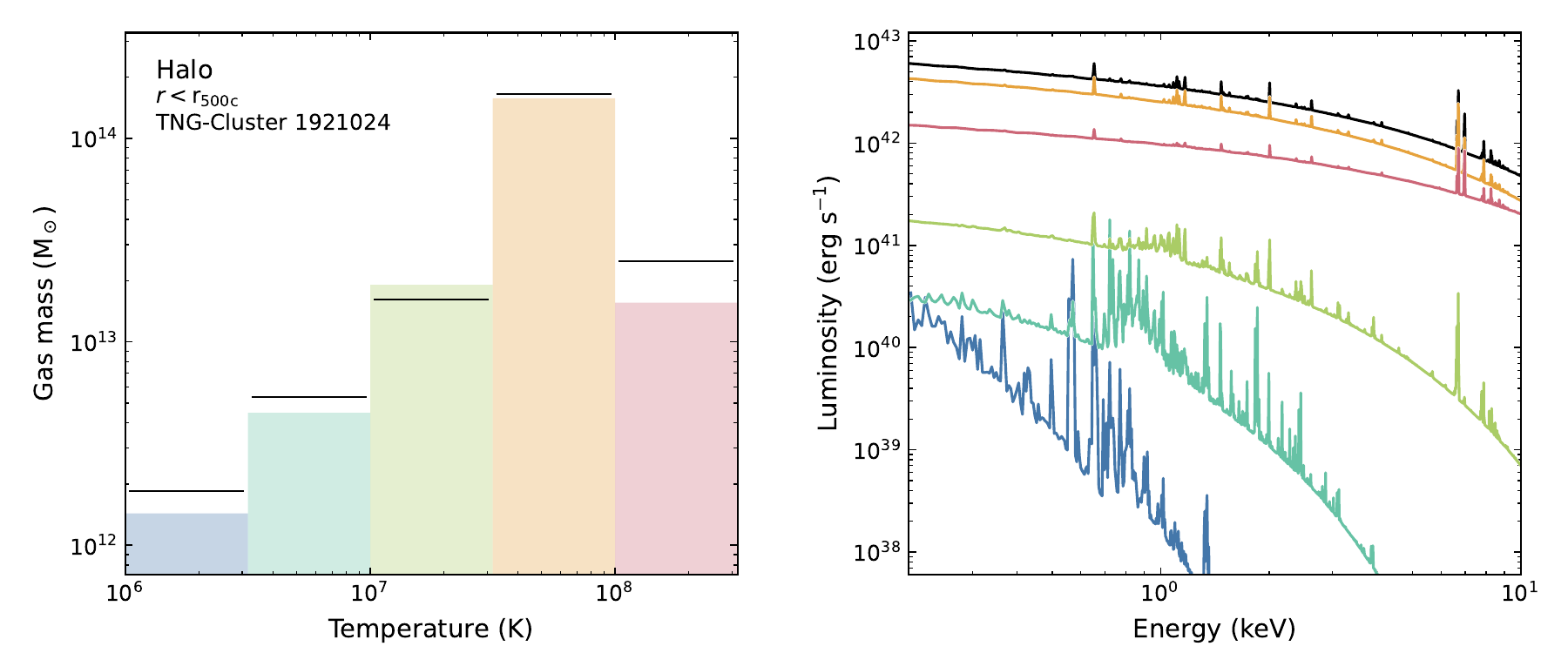}
    \includegraphics[width = \linewidth]{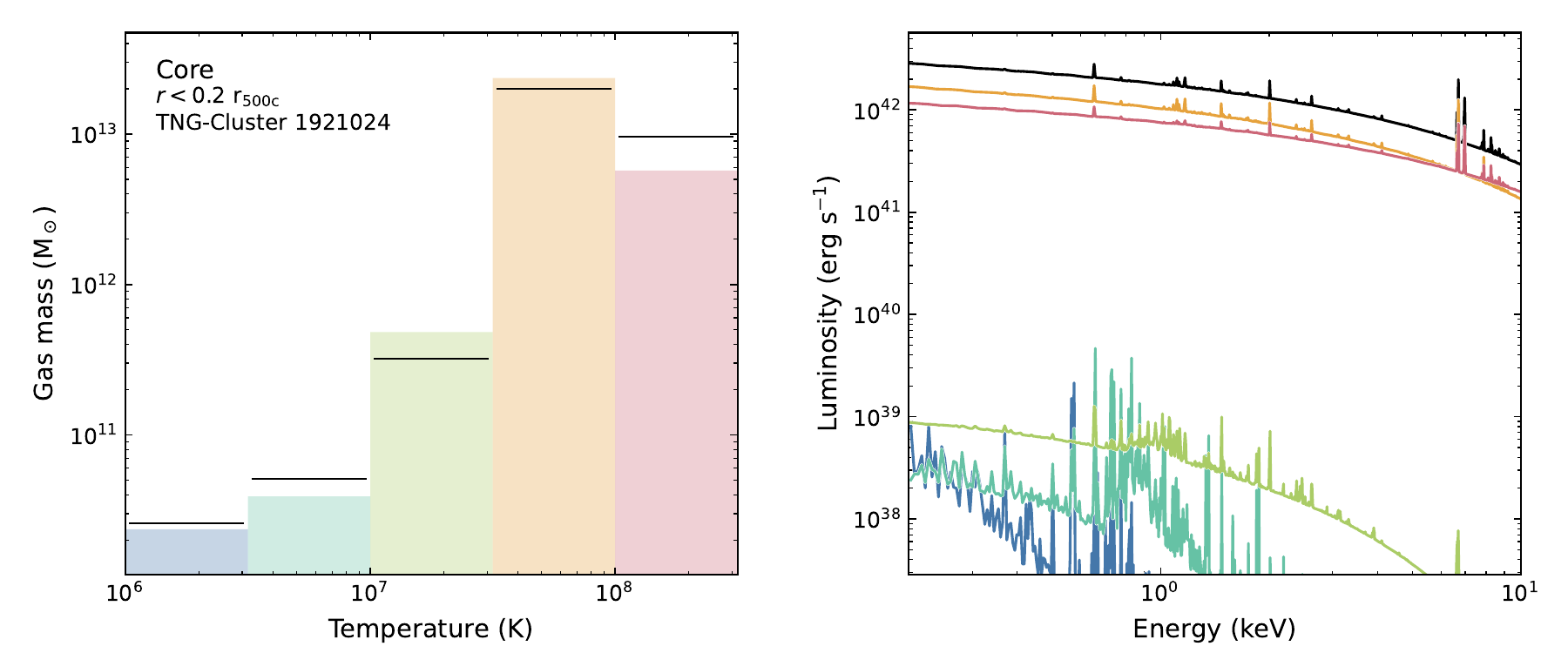}
    \includegraphics[width = \linewidth]{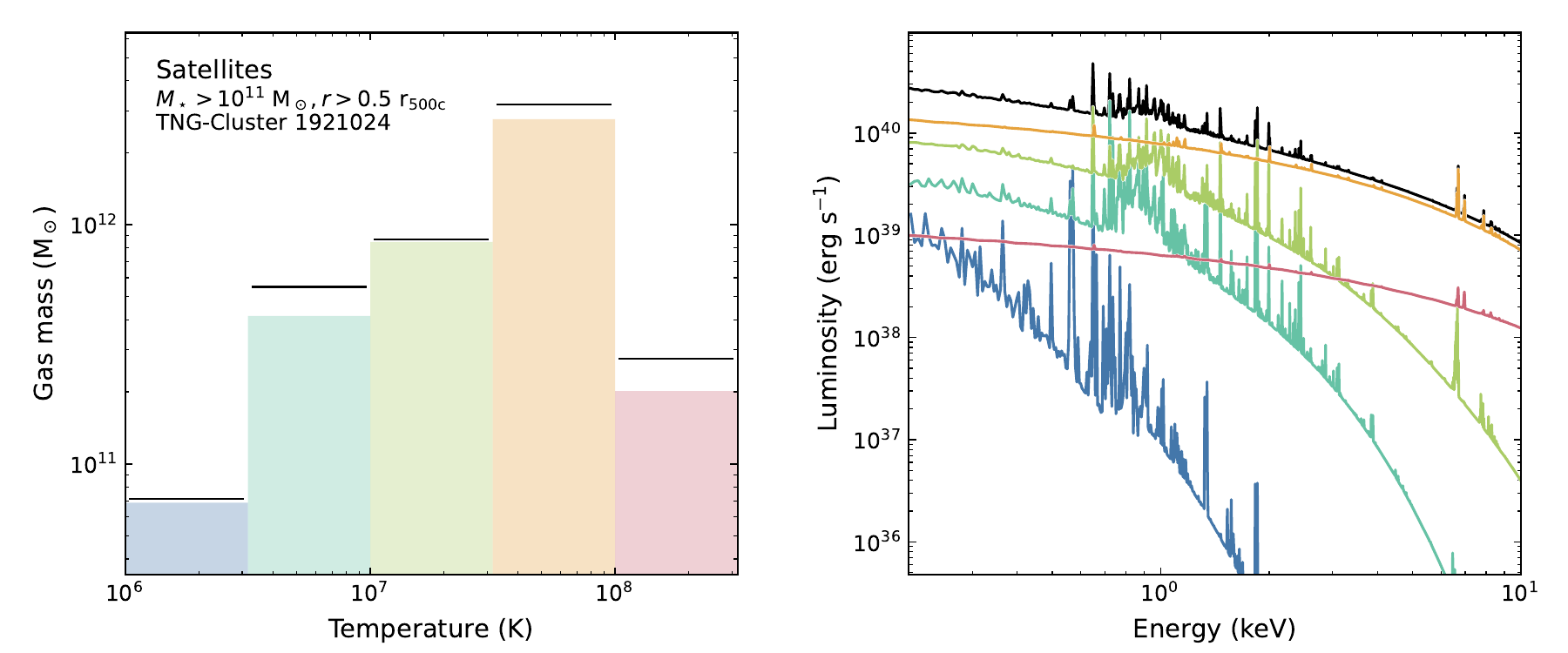}
    
    \caption{Temperature distributions and predictions for an example $M_{\rm 500c} = 10^{15}~\mathrm{M_\odot}$ cluster, showing the summed result for the full $R_{\rm 500c}$ region (316 sightlines) on top, the core region ($<0.2 ~ R_{\rm 500c}$, 12 sightlines) in the middle, and massive satellites in the outskirts ($M_\star > 10^{11}~\mathrm{M_\odot}$; $r > 0.5~R_{\rm 500c}$, 12 sightlines) in the bottom panels. Black horizontal lines indicate the true gas mass in each bin, coloured bars indicate the predicted gas mass in the same bin. Right panels show the total spectrum in black: this is the only input to the neural network. Coloured spectra show, for each temperature bin in the left panel, the corresponding spectrum (which the network never sees). This shows that the cooler gas is sub-dominant in the total spectrum by multiple orders of magnitude. Nevertheless, we can still accurately predict the gas mass, likely due to the subtle changes in individual emission lines.}
    \label{fig:sightline_and_spectra}
\end{figure*}

Having ascertained accurate predictions for individual sightlines, mass-weighted temperatures of entire clusters, and the spatial coherence of these predictions on cluster maps, we now continue on the cluster-scale and investigate the contribution of different temperature components to the predictions. In Figure \ref{fig:sightline_and_spectra} we show, top to bottom, results for the a $10^{15}~\mathrm{M_\odot}$ halo ($r \leq R_{\rm 500c}$), its core region ($r \leq 0.2 ~ R_{\rm 500c}$), and the sightlines centred on massive satellites in the outskirts ($M_\star > 10^{11}~\mathrm{M_\odot}$; $r > 0.5~R_{\rm 500c}$). The latter are chosen to capture satellites before they are fully stripped, and we expect those to harbour more cool gas compared the surrounding ICM. For the core region, we expect an increase in the hot gas fraction. In these figures the black lines represent the true gas mass in each temperature bin, and the coloured bars the predicted gas mass. The colours of the temperature bins are the same in the histograms and for the spectra.

Comparing the black line histograms in the left panels, we see the expected trends between the three different regions. The core contains more hot gas compared to the total ICM, and massive satellites harbour more cooler ($T < 10^7 ~\mathrm{K}$), and less hot gas. The predictions follow this trend. The prediction for both hot and cooler gas is equally accurate in all regions.

In the right panels, the summed spectra from all the sightlines in each region are shown. They are split by temperature bin, using the same colour as the left panels. In all regions, the virial temperature gas (for this halo $9\times10^7 ~\mathrm{K}$, orange) dominates. For the halo-wide and core region, this is followed by the super-virial gas. In the core the super-virial gas contributes almost equally to the virial gas phase, while it is $3\times$ subdominant on the entire halo scale. The emission from the cool gas ($T < 10^7 ~\mathrm{K}$) makes up $\approx 0.5\%$ of the entire halo, $\approx 0.1\%$ of the core, but reaches partition with the virial gas phase in the sightlines centered on massive satellites. For these latter sightlines we see a strong line complex around $1~\mathrm{keV}$ arising from the cooler gas. 

Despite the fact that in an absolute sense all the regions are dominated by virial, or near-virial temperature gas, our model is able to succesfully recover the change in cool gas fraction between the different regions. We emphasise again that the neural network only sees the total spectrum (black line), making it even more impressive that the sub-dominant components are recovered. While the subdominant component is the cooler gas for the halo-wide and core sightline, in the satellite sightlines the super-virial gas is subdominant by at least and order of magnitude, despite that, the prediction is still as accurate as for the regions where it dominates.

We will study the importance of different parts of the spectrum to the predictions in more detail in Section \ref{sec:model_explanation}.

\begin{figure*}
    \centering
    \includegraphics[width=\linewidth]{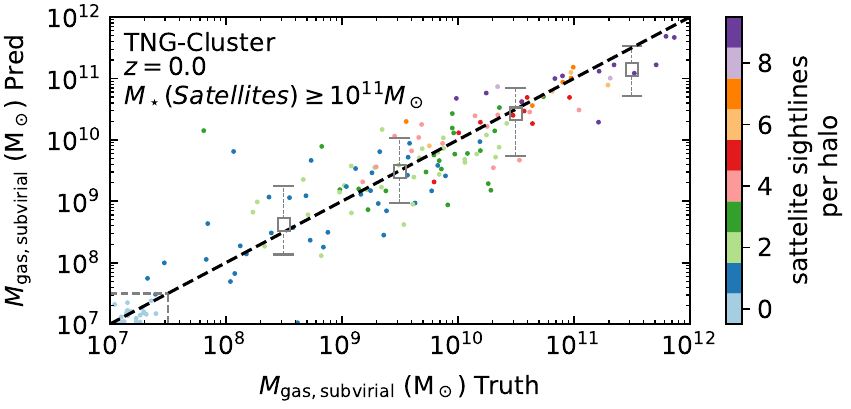}
    \caption{\textbf{Performance of our method on the sub-virial gas in satellites.} We show the total sub-virial gas mass in lines-of-sight intersecting with satellite galaxies of $M_{\star} \geq 10^{11}~\mathrm{M_\odot}$. Sub-virial gas is defined as all gas at least $0.5~\mathrm{dex}$ colder than the mass-weighted temperature of the halo. Each scatter point represents one halo, its color indicates how many lines-of-sight satisfying the satellite mass criterion there are. Haloes with no satellites satisfying this criterion are randomly assigned sub-virial gas mass values between $10^{7-7.4}~\mathrm{M_\odot}$, contained in the box in the lower left corner. In 4 bins of $1~\mathrm{dex}$ in mass we calculate the median and 16th and 84th percentiles of the prediction, this is shown with grey squares and error bars. The sub-virial gas mass along lines-of-sights of massive satellites is well predicted, with the prediction improving for larger sub-virial gas masses, and if there are multiple massive satellites present. The latter could be due to the averaging of multiple lines-of-sight, yielding a more accurate average prediction.}
    \label{fig:cold-gas-mass-satellites}
    
    \centering
    \includegraphics[width=.8\linewidth]{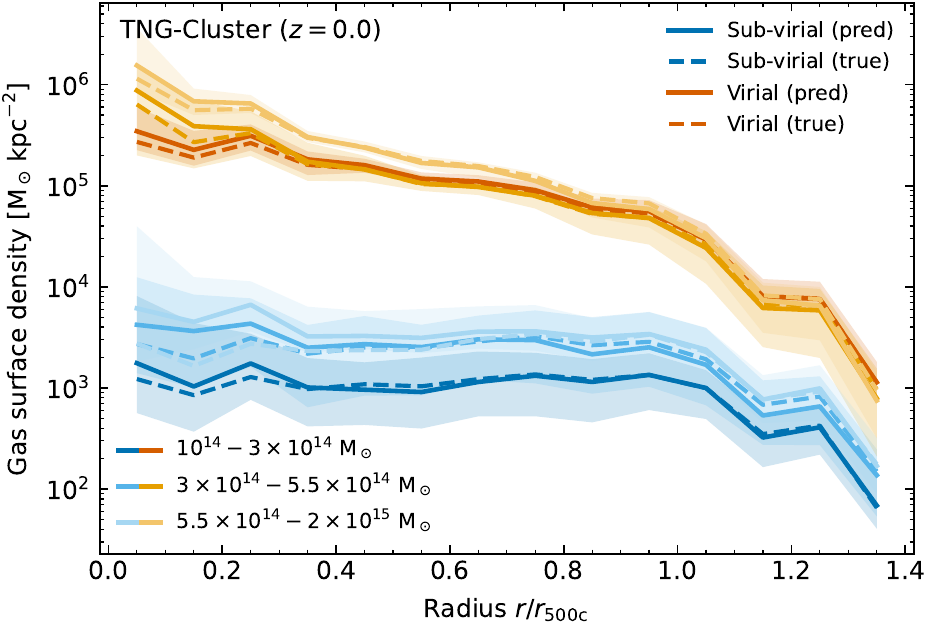}
    \caption{\textbf{Performance of our method on the sub-virial gas profiles.} Median surface density profiles of virial temperature gas (red), and sub-virial gas (blue) in 3 halo mass bins. Solid lines and shaded regions show the inference from with our model, and halo-to-halo variation. Dashed lines show the true profiles, both are measured from the pixelated maps. Both the virial gas and the cooler gas profile are recovered excellently at all halo masses. The virial gas follows the halo density profile, decreasing monotonically with radius, whereas the cooler gas has a near constant surface density within $R_{\rm 500c}$.}
    \label{fig:gas_profiles}
\end{figure*}
\subsection{Sub-virial gas in satellites} \label{sec:sattelitle_predicitons}
As a further test of our method on recovering cool gas across a range of scenarios, we apply it to lines-of-sight containing massive satellites in the cluster outskirts. As in the previous section, we pick sightlines with $M_{\star} > 10^{11}~\mathrm{M_\odot}$ and $r_{\rm sightline} > 0.5 ~ R_{\rm 500c}$, where the stellar mass refers to the most-massive satellite with its centre of mass within the line-of-sight. Such massive satellites at large cluster-centric radii are expected to retain a large amount of cooler, sub-virial gas, as they have not yet been fully ram-pressure stripped \citep[e.g.][]{rohr_jellyfish_2023}.
To quantify the amount of sub-virial gas in each sightline, we define 
\begin{equation}
    M_{\rm gas, sub-virial}  = M_{\rm gas}(\log T < \log T_{\rm MW,halo} - 0.5),
     \label{eq:cold_gas}
\end{equation}
or the gas that is at least $0.5 ~\mathrm{dex}$ colder than the mass-weighted temperature of the halo.

In Figure \ref{fig:cold-gas-mass-satellites} we show the result of all clusters in the test set. Each cluster is represented by one point, that indicates the total mass of sub-virial gas in all sightlines satisfying the above criteria. The colour indicates the number of such sightlines. A small number of, mostly low-mass, clusters have no sightlines with massive satellites, these are depicted in the lower left corner with a light-blue color, and a random small value for sub-virial gas mass.
For the clusters that do contain massive satellites, our method reproduces the sub-virial gas mass in the sightlines quite well. Though there is considerable scatter, the predictions follow the 1-to-1 line across 4 orders of magnitude of sub-virial gas mass. The mass correlates with the number of satellites, but even within a fixed number of satellites (same colour), we see a clear correlation between sub-virial gas mass prediction and true value. Furthermore, there is considerable overlap among colours, showing that some clusters have more satellites with less sub-virial gas, compared to clusters with fewer satellites but harbouring more gas in that phase.
In grey markers and error bars we show the median and 16th-84th percentile of the prediction in bins of $1~\mathrm{dex}$ between $M_{\rm gas,sub-virial}=10^{8}-10^{12}~\mathrm{M_\odot}$.

Overall, this result shows the robustness of our method in predicting the sub-virial gas phase across a diverse range of clusters. This is a clear proof that our method can extract the complex multi-phase thermodynamics of clusters, even away from the smooth ICM and close to satellites.

\subsection{Multi-temperature gas profiles} \label{sec:cold_gas_profiles}

The next question that arises is whether our model can successfully return even the 1D radial distribution of different gas phases? We now study the radial distribution of virial temperature gas as well as sub-virial gas. Virial temperature gas is defined as all the gas that falls into the same temperature bin as the mass-weighted temperature of the halo. Sub-virial gas is the sum of all gas which is at least $0.5~\mathrm{dex}$ cooler, which is the same definition as in previous sections and Equation \ref{eq:cold_gas}.
We divide the pixelated maps of each cluster into 14 radial bins spanning from 0 - 1.4 $R_{\rm 500c}$, each bin being exactly $0.1~\mathrm{R_{\rm 500c}}$ wide. This results in varying number of sightlines in each radial bin, as the radial bin surface increases with radius.

In Figure \ref{fig:gas_profiles} we show the radial gas surface density profiles of these two phases, split into three halo mass bins with equal numbers of haloes. The shaded area shows the 16th to 84th percentile of halo-to-halo variation. We observe that both the virial gas surface density profile, as well as the sub-virial gas profiles are in excellent agreement with the ground truth at all halo masses. This indicates an unbiased inference of both gas phases. Furthermore, we see that where the virial temperature gas follows the monotonically decreasing halo density profile, as it represents the bulk of the gas, the sub-virial surface density is nearly constant with radius. In the core, the virial gas dominates by up to 3 orders of magnitude, decreasing to approximately 1 order of magnitude at $R_{\rm 500c}$. 

This further consolidates the veracity of our predictions: not only can we recover the prominent cooler gas in satellites, we can capture the highly subdominant cooler gas in the diffuse ICM across the entire cluster.
\begin{figure}
    \centering
    \includegraphics[width=\linewidth]{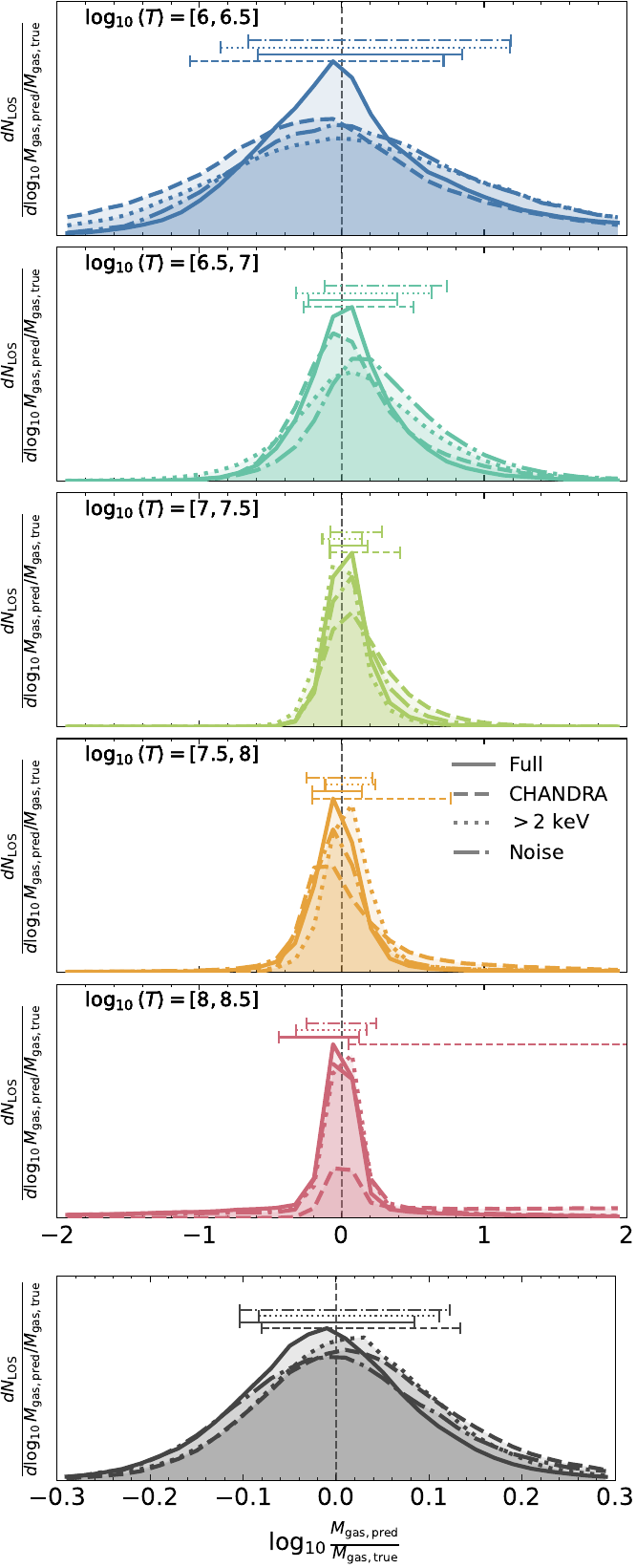}
    \caption{PDFs of the prediction error of the gas mass in the 5 temperature bins, and the total gas mass, for varying properties of the input X-ray spectra. Solid lines show the results using the full spectrum, for the dotted lines we limit it to use only $E>~\mathrm{2keV}$ (to mimic the case of the gate-valve issue with \textit{XRISM}), for the dashed lines the spectrum is convolved with the \textit{Chandra} response, and for the dash-dotted line we have applied realistic noise. The error bars show the 16th-84th percentile ranges for each of these different models. We note that all permutations have very similar quality predictions. Only for the highest temperature bin does \textit{Chandra} result in a tail of over-estimations.}
    \label{fig:CHANDRA_gatevalve}
\end{figure}
\subsection{Degrading the spectra} \label{sec:spectral_resolution}
We now assess the ability of our network architecture to capture changes in spectral resolution and bandpass. Two scenarios are explored (see Section \ref{sec:instrument_response})
\begin{itemize}
    \item[1.] Applying a bandpass $E > 2~\mathrm{keV}$ to mimic the closed gate valve on the \textit{XRISM} telescope
    \item[2.] A convolution with the \textit{Chandra} response function, which reduces the spectral resolution and smoothes emission lines
    \item[3.] Adding photon noise to the \textit{XRISM}-like spectra
\end{itemize}

For all cases we regenerate the training data and retrain the network. The architecture is kept fixed, but the hyper-parameters are free to vary. We do not find consistent changes in the preferred hyper-parameters.
In Figure \ref{fig:CHANDRA_gatevalve} we visualise the accuracy of the prediction in these three cases, all removing information in different ways. Each of the first 5 panels corresponds to a temperature bin, the final panel corresponds to the gas mass prediction. Solid lines show the default results which we have explored extensively in this work. Only utilizing the part of the spectrum $E>2~\mathrm{keV}$ is shown by the dotted line, the convolution with the \textit{Chandra} response by the dashed line, and the noise model with the dash-dotted line. The differences are surprisingly small. Only the highest temperature bin, which already showed the worst predictions for the full spectrum, differs significantly. Convolving with the \textit{Chandra} response leads to a consistent over-prediction, this might be due to the low sensitivity of \textit{Chandra} to highest energies.

The remarkable consistency in the quality of the inference shows that our methodology is sufficiently powerful to adapt to changing inputs, as long as the network is retrained.

\section{Model explanation and interpretability} \label{sec:model_explanation}
We now delve into the internal mechanics of our neural network and aim to explain which features in the input spectra contribute to the final predictions. We approach this in several steps. First in Section \ref{sec:feature_importance} we apply a SHAP analysis to highlight the relevant and informative parts of the input spectra. In Section \ref{sec:ablation_cnn} we then look at the internal pathways of our neural network and their relative contributions to the predictions. Finally, in Section \ref{sec:importance_line_regions} we select three prominent emission line regions, which are often used in observations, and study how removing those lines changes the predictions.

\subsection{Feature importance} \label{sec:feature_importance}
\begin{figure*}
    \centering
    \includegraphics[width=\linewidth]{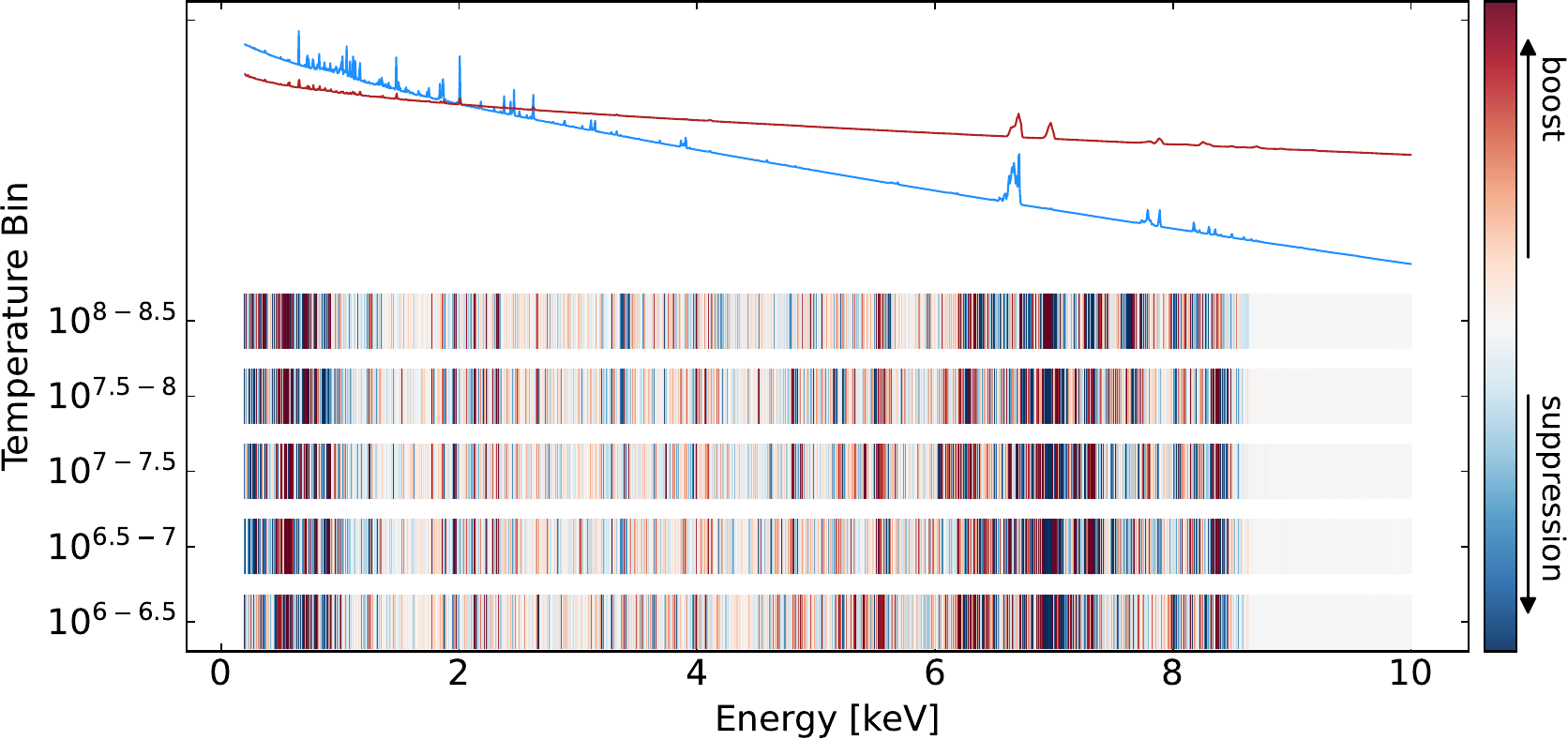}
    \caption{Feature importance of the input X-ray spectra for the prediction of different temperatures. Red and blue colours respectively indicate the top 5\% positive and negative SHAP values for predicting the gas mass in each temperature bin. A positive value (red) indicates that a larger flux at that energy boosts the prediction, while for a negative value (blue) a larger flux would suppress the prediction. We note that there are two regions of primary interest to the network: the first is low energies ($E < 1~\mathrm{keV}$), the second at high energies ($6 ~\mathrm{keV} < E < 8 ~\mathrm{keV}$), with other regions of the spectrum contributing little information. Two example spectra for a cool sightline (blue) and hot sightline (red) are shown on top.}
    \label{fig:shap_values}
\end{figure*}
Which parts of the X-ray spectrum is informative to our neural network approach? Do these align with the regions that are usually considered when fitting for temperatures from X-ray observations? We make use of SHAP values to interpret which part of the inputs are up- or down-weighting the prediction \citep{lundberg_unified_2017}. SHAP values originate in game theory and are able to capture the non-linear relationship between different input features, by comparing their relative contributions with respect to a baseline. In our case, the baseline is the average of the entire test data set. We then calculate SHAP values for a small, uniformly sampled subset of the test set. 

In Figure \ref{fig:shap_values} we show, per energy bin, the SHAP value for each target temperature bin. That means, for each temperature bin we can see which parts of the spectra contribute positively (red) or negatively (blue). For clarity, we truncate the color scale at the 5\% largest positive and negative values, such that we can also see the more subtle contributions. Two reference example spectra are shown on top: one spectrum for a cooler sightline in blue, and one for a hotter sightline in red. However, as for each of our sightline, there is a mixture of cooler and hotter gas present, giving rise to a complex spectrum that is a super-position of multiple phases.

Though the exact location of positive and negative contributions is seemingly quite chaotic, we see that it correlates with two important line-rich regions. First, there are large contributions to the predictions below $1 ~\mathrm{keV}$ where many low-temperature emission lines are known to be present. Secondly, there is a region between $6 - 8 ~\mathrm{keV}$ with a strong iron-line complex, which dominates the spectrum. We note that both the low- and high-temperature predictions depend on both of these regions.

In Appendix \ref{sec:appendix_feature_importance_chandra} we repeat this for spectra convolved with the \textit{Chandra} response and find broadly the same result, though with notably more importance to the shape of the spectrum and less to the emission line regions.

\subsection{Ablation of Neural Network pathways}\label{sec:ablation_cnn}
We perform a mean ablation study in order to quantify the importance of the two pathways in our network: The stride-heavy convolutional block focussing on local information, and the pooling-heavy block focussing on more global information (see Figure \ref{fig:nn_architecture} for reference). 

Both of these NN pathways produce a latent vector. We can hence compute the mean of those two latent vectors over the entire test set, and generate predictions while keeping one of them fixed to the mean value. This  essentially removes any predictive power coming from that pathway. The importance of each pathway is computed as
\begin{align}
    R_1 &= \mathcal{M} - \mathcal{M}(\text{mean pathway 1)} \, ,\\
    R_2 &= \mathcal{M} - \mathcal{M}(\text{mean pathway 2)} \, .
\end{align}
The non-linear interaction term between the two pathways is given by
\begin{align} 
\begin{split}
    R_I =  \mathcal{M} &- \mathcal{M}(\text{mean pathway 2)} \\ &- \mathcal{M}(\text{mean pathway 1)} \\&+ \mathcal{M}(\text{mean pathway 1, mean pathway 2)} \, ,
\end{split}
\end{align}
then their relative importances are given as
\begin{align}
    S_i = \frac{|R_i|}{|R_1|+|R_2| + |R_I|} \, .
\end{align}
Using these equations, we find that the pathway focussing on narrow features, i.e. spectral lines, dominates slightly over the pathway focussing on broad features, i.e. the continuum shape. Namely, the relative importances are $S_{\text{narrow}} = 0.56$, $S_{\text{broad}} = 0.34$.
Whereas the non-linear interaction between the two pathways is comparatively small ($S_I = 0.10$).

\subsection{The importance of specific emission-line regions}\label{sec:importance_line_regions}
We have seen that there are two major regions of importance to the predictions, low-energy lines ($E < 1~\mathrm{keV}$) and high-energy lines ($6~\mathrm{keV} < E < 8~\mathrm{keV}$). The latter region corresponds to two strong line complexes, namely FeXXV and FeXXVI, whereas the first region is an amalgamation of many emission lines originating in cooler gas. We have also seen that the NN pathway focussing on local information, and hence likely to pick up information about emission lines, slightly dominates over the global-information pathway. 
We know investigate whether specific line regions have a strong individual contribution to the prediction of each temperature bin. We focus on the lines mentioned above, FeXXV, FeXXVI and the full region below $1~\mathrm{keV}$.

To study the effect of those regions, we remove each of those three line regions one-by-one, or in combinations, by replacing them with the continuum.
We then measure the ratio between the mean absolute prediction error by removing those lines and when keeping them. A larger ratio would indicate that removing them made the prediction worse, and hence they have a strong individual contribution.

In Table \ref{tab:line_importance} we show these ratios for each of the temperature bins and for the total gas mass. 
Going through the line (regions) one by one:
\paragraph*{FeXXV:} Removing this emission line leads to worse predictions for all gas above $10^{7}~\mathrm{K}$, and for the hottest gas phase the prediction error even doubles. We ascribe the fact that the error on the $10^{7.5-8}~\mathrm{K}$ gas is smaller than both the cooler and hotter gas to non-linear interactions between different parts of the input spectrum.
\paragraph*{FeXXVI:} Removing this emission line makes the prediction of just the two highest temperature bins worse ($>10^{7.5}~\mathrm{K}$), by respectively 100 per cent and 75 per cent. 

This shows the sensitivity of FeXXVI to the very highest temperature gas, and indicates that our network makes use of this domain specific knowledge, without ever being explicitly informed about it.
\paragraph*{E < $1~\mathrm{keV}$:} Removing all the low energy emission lines makes the prediction worse by 30-60 per cent for all gas with $<10^{8.0}~\mathrm{K}$. This effect is strongest, but not limited to, the temperature ranges $10^{6.5-7}~\mathrm{K}$. The susceptibility to such a wide temperature range is likely due to the large number of emission lines present in this region, and the large extent of the region in energy. This would lead to many non-linear interaction not captured by the simple removal of all emission lines.
\paragraph*{FeXXV+FeXXVI:} Removing both these emission lines at the same time shows that there is significant interaction between their features. The prediction on the $10^{7.5-8}~\mathrm{K}$ gas remains twice as bad as the baseline, but the gas at $10^{8-8.5}~\mathrm{K}$ is curiously better predicted than when removing only one of the lines. One interpretation would be that the ratio between these two emission line complexes is important, and removing both of them changes the ratio by less (it would be the ratio of the continuum at the two line positions), than removing only one emission line.
\paragraph*{Combined:} Removing all three emission line regions shows a significant drop in predicitive accuracy for all our output quantities. At this point, most of the emission line information that the model focusses on (per Fig. \ref{fig:shap_values}) has been removed, hence this is not surprising. The combined removal of these 3 regions also makes the prediction on the total gas mass worse by 65\%, showing that the gas mass prediction is sensitive to emission lines, and not just the integrated luminosity.

\begin{table}
\centering
\caption{Relative change in mean absolute error when removing one or more line regions from the input spectrum for the 5 temperature bins and the total gas mass.}
\label{tab:total_errors}
\begin{tabular}{lccccc}
\hline
Temp bin & FeXXV & FeXXVI & $E < 1~\mathrm{keV}$ & FeXXV  & Combined \\
($\log_{10} \mathrm{K})$& & & & +FeXXVI & \\
\hline
6.0--6.5      & 1.026 & 1.025 & 1.287 & 1.022  & 1.293 \\
6.5--7.0      & 1.040 & 1.016 & 1.635 & 0.985  & 1.664 \\
7.0--7.5      & 1.546 & 1.072 & 1.375 & 1.166  & 1.627 \\
7.5--8.0      & 1.287 & 1.991 & 1.380 & 2.003  & 2.460 \\
8.0--8.5      & 2.054 & 1.762 & 1.023 & 1.350  & 1.157 \\
$M_{\rm gas}$ & 1.211 & 1.104 & 1.201 & 1.350  & 1.652 \\
\hline
\end{tabular} \label{tab:line_importance}
\end{table}

\section{Discussion} \label{sec:discussion}

\subsection{Alternative approaches to uncovering mutli-temperature gas}
Recent \textit{XRISM} observations have found tentative evidence of multi-temperature gas. To unearth this gas, spectral fits are performed on a wide spectral energy range, which is subdivided into regions known to contain emission lines tracing different ionisation temperatures. At the moment, such efforts are able to uncover two temperature gas, with the two phases typically separated by a factor 2 in temperature \citep{majumder_spectrally_2026, simionescu_dynamics_2026}. This represents a very narrow temperature range compared to the $2.5~\mathrm{dex}$ in temperature considered in this work. Further progress is limited by the need of very high signal-to-noise for the spectral line fits to succeed, currently barely attainable with \textit{XRISM}. Fundamentally, with the gate-valve closed, and the telescope response function limited to $E > 2~\mathrm{keV}$, such studies are limited to high ionisation temperature lines and unable to probe sub-keV gas. Looking at adjacent regions, instead of co-spatial, \textit{XRISM} does find clear temperature gradients, hinting at cooling and mixing happening in the ICM, which would give rise to multi-temperature distributions \citep{sarkar_xrism_2025}.
A second method, using line ratios rather than a direct spectral fit, seems similarly limited \citep{meunier_probable_2026}.
With the advent of \textit{NewAthena}, the use of line ratios will likely increase drastically and provide valuable insight into the gas thermodynamics. However, it fails to incorporate the shape of the continuum, where subtle deviations can indicate sub- or super-virial gas phases. A method like ours, which simultaneously takes in the spectral lines and the continuum, is arguably better suited to an unbiased extraction of the gas temperature distribution.

\subsection{Sensitivity to resolution and noise}
Ideally, X-ray observations would give access to spatially resolved, multi-temperature distributions. This would allow the accurate assessment of the thermodynamic state of clusters, give insight in the entrainment of cold gas, the distribution of shock fronts, and filaments. The current observational landscape, combined with the current inference tools, however, does not allow for such precision. High spatial resolution observations, such as with \textit{Chandra}, do not have the spectral resolution to get accurate temperature measurements with standard fitting methods, let alone multi-temperature inference. On the other end, new \textit{XRISM} observations are starting to open the road to multi-temperature inference, albeit with caveats discussed in the previous section, but with significantly worse spatial resolution.

We have shown that our method can go beyond that and allows a robust inference of multi-temperature gas across $2.5~\mathrm{dex}$ in temperature, independent of spectral resolution (see Section \ref{sec:sightline_performance} and \ref{sec:spectral_resolution}). Though we refer the exploration of realistic telescope mocks to future work, the relative insensitivity to spectral resolution and noise suggests that accurate predictions are possible even in the absence of high-information signal. We attribute this to the retraining of the network. When using only energies above the gate-valve cut-off, convolving with the \textit{Chandra} energy response function, or adding realistic levels of noise, we retrain the network. In practice, this means that different information is used in each case. For example, by inspecting the network weights, we see that after convolution with the \textit{Chandra} response, the network places much more importance on the shape of smoothed-out line regions, rather than the heights of individual lines.
Such a flexible approach, optimised for a certain set of observations, shows tremendous promise and gives us confidence that our results will be robust even when introducing realistic observational noise and artefacts.

\subsection{Generalization to other simulations, i.e Flamingo}
\begin{figure}
    \centering
    \includegraphics[width=\linewidth]{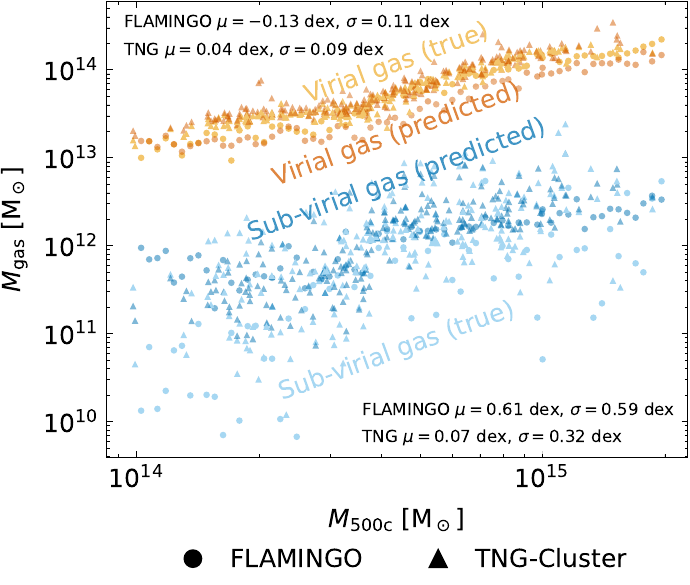}
    \caption{\textbf{Generalization to other galaxy cluster simulation models.} Predicted and true gas mass as a function of halo mass for the virial and sub-virial temperature component in TNG-Cluster and with the same model applied to FLAMINGO. The values in the top-left and bottom-right corner indicate the mean error and scatter around the mean error for the virial and sub-virial phase. We see that the FLAMINGO predictions closely track the true values for the virial gas phase, but result in an over-prediction for the sub-virial gas.}
    \label{fig:flamingo}
\end{figure}

To quantify how well our model generalizes to other simulation models, we generate spectra and temperature distributions from the fiducial FLAMINGO simulation, following the same procedure as for TNG-Cluster. From FLAMINGO we select 64 clusters equally spaced in mass between $M_{\rm 500c} = 10^{14.0} - 10^{15.3}~\mathrm{M_\odot}$ for an identical halo mass range as covered by TNG-Cluster. We then apply our neural network model, which is trained on TNG-Cluster, to the spectra extracted from FLAMINGO. The results are shown in Fig. \ref{fig:flamingo}. 

The results for the virial gas show that both simulations largely overlap in the covered parameter space. There is a clear halo mass trend with limited scatter. The model (trained on TNG) works extremely well for TNG, resulting in a negligble bias of $0.04~\mathrm{dex}$ with a scatter of $0.09~\mathrm{dex}$. On FLAMINGO the bias and scatter are slightly larger at $-0.13~\mathrm{dex}$ and $0.09~\mathrm{dex}$, but still well within typical observational error bars. This shows that for the virial gas phase, our model generalizes well to the FLAMINGO simulations.

For the sub-virial gas we see a separation in the true values. TNG-Cluster haloes typically contain more of this gas compared to FLAMINGO. In fact, we find that many FLAMINGO sightlines contain none of the coldest temperature gas we consider ($10^{6-6.5}~\mathrm{K}$). This could very well be a resolution effect, the average gas mass in the coldest temperature bin for TNG-Cluster is less than the mass resolution of FLAMINGO.
Nonetheless, the average trend is still reproduced in FLAMINGO albeit with a bias of $0.61~\mathrm{dex}$. We note however that current observational methods are entirely unable to constrain the sub-virial gas, especially when it is 2 orders of magnitudes less abundant than the virial phase. Hence, a constraint on the sub-virial is a significant improvement on current methods, even with the bias we are seeing for FLAMINGO.

In general, these results call for a greater emphasis on the generalization performance across resolution. This will be a focus of future work.

\section{Conclusions} \label{sec:concolusions}
We have presented a direct inference method for the true mass-weighted temperature distribution of a line-of-sight of diffuse X-ray emitting gas in galaxy clusters from both low and high resolution X-ray spectra (0.2-10 keV). Our method is based on the principles of Simulation-Based-Inference (SBI), using an optimised convolutional neural network trained on sightlines generated from all the 352 $z=0$ clusters from the cosmological MHD galaxy formation simulation TNG-Cluster.
We predict the temperature distribution in 5 bins of $0.5~\mathrm{dex}$ between $10^6 - 10^{8.5}~\mathrm{K}$, as well as the total gas mass ($M_{\rm gas}$) along each sightline. Each cluster is divided into 400 sightlines on a 20x20 grid, with each sightline extending $0.1 \times 0.1 ~\mathrm{R_{\rm 500c}}$.
The training is done on 46,400 sightlines (116 clusters), while the test set consists of 93,600 sightlines (234 clusters). This assures a large variety in X-ray spectra and temperature distributions.

Our main findings are:
\begin{itemize}
    \item The best performing neural network consists of two separate channels: One channel to treat global information, such as the continuum, and a second channel for local information, such a spectral lines. The normalisation, i.e. X-ray luminosity, is passed as an additional input after the shape of the spectrum has been processed. This additional input needs embedding, but can otherwise be flexibly extended to other scalar inputs (Figure \ref{fig:nn_architecture}).
    \item Our method (based on individual sightlines) can infer mass-weighted temperatures for entire clusters with an average prediction error of $-0.009 \pm 0.034 ~\mathrm{dex}$, and the total gas mass within $0.043 \pm 0.078 ~\mathrm{dex}$, across the halo masses ranging from $M_{\rm 500c} = 10^{14.3} - 10^{15.5} ~\mathrm{M_\odot}$ (Figure \ref{fig:halo_predictions}).
    \item For the 5 temperature bins, on a sightline basis, the inference of the temperature distribution ($10^{6 - 8.5}~\mathrm{K}$) is accurate and unbiased. For lower temperatures ($< 10^{7} ~\mathrm{K}$), which usually contribute minimally to the spectrum, the scatter in the inference error increases, though the majority of sightlines are still well predicted. We find that the gas mass in a given temperature bin can be accurately predicted as long as the fractional contribution of gas at that temperature to the spectrum is at least $10^{-4}$ (Figures \ref{fig:error_per_t_bin} and \ref{fig:mse_neighbours_network}).
    \item Putting together the independent predictions for all sightlines of each cluster, we can obtain the temperature maps of clusters and their surrounding gas. We accurately recover hot plumes, cold fronts, merger shocks and more peculiar features. This shows that the inference is of sufficiently high quality to reconstruct the 2D temperature maps of clusters (Figure \ref{fig:predicted_true_maps}).
    \item We study the sub-virial gas of clusters, defined as gas that is at least $0.5~\mathrm{dex}$ below the mass-weighted temperature. We recover cold gas in sightlines centered on cluster cores (Figure \ref{fig:sightline_and_spectra}), massive satellites (Figure \ref{fig:cold-gas-mass-satellites}), and the cold gas radial surface density profile (Figure \ref{fig:gas_profiles}). In the profiles, gas at the virial temperature shows a monotonically decreasing surface density with radius, whereas the sub-virial gas has a constant surface density within $R_{\rm 500c}$. 
    \item Feature importance studies reveal that two regimes in the spectrum contribute most to the temperature predictions: lines originating from cooler gas ($E<1~\mathrm{keV}$), and lines from hotter gas $6~\mathrm{keV} < E < 8~\mathrm{keV}$ (Figure \ref{fig:shap_values}). We also find that narrow features in the spectrum contribute around 55 per cent to the predictive power of the network, compared with 35 per cent for broad features, and 10 per cent for the non-linear interplay between both. Individual emission lines contribute to the prediction in complex non-linear ways.
    \item We confirm the generalizability of our method to other simulation models. In particular we test our model, trained on TNG-Cluster, on clusters extracted from the FLAMINGO simulation and find good performance for virial temperature gas mass with an average prediction error of $0.1~\mathrm{dex}$. Fot the sub-virial gas mass the prediction error is on average $0.6~\mathrm{dex}$. (Figure \ref{fig:flamingo}).
\\
\\
Accurate inference of the temperature distribution of galaxy groups and clusters has been a long standing problem. For a long time the limit was set by low spectral resolution. However, even with the advent of \textit{XRISM} and its high spectral resolution micro-calorimeter, temperature inference remains fraught with biases. In this work we have shown a path forward to obtain not only accurate, mass-weighted, average temperatures, but also a sense of the underlying temperature distribution. 
In follow-up work we will use mock observations of both low and high spectral resolution telescopes, including realistic noise and instrumental effects. Given the relative flexibility of our method shown by its ability to work, and recover cool gas phases, both in a restricted spectral range (like that of \textit{XRISM} with the gate-valve closed) and low spectral resolution (after convolution with the \textit{Chandra} response), we expect to still be able to accurately recover the temperature distribution. Future work will build on our methodology to extract not only the temperature, but also the metallicity and potentially the velocity distribution of the gas along the line-of-sight. More work will also have to be done to study the generalizability to different numerical resolutions and simulation models, though the initial results are very promising.
Given these results and prospects, we see a bright future for the use of simulation based inference methods in X-ray astronomy.
    
\end{itemize}

\section*{Acknowledgements}
We thank Aarya Patil for useful discussions at the inception of this project.
JB, AP, MP, and DC acknowledge funding from the European Union (ERC, COSMIC-KEY, 101087822, PI: Pillepich). DC is a fellow of the International Max Planck Research School for Astronomy and Cosmic Physics at the University of Heidelberg (IMPRS-HD). DN acknowledges funding from the Deutsche Forschungsgemeinschaft (DFG) through an Emmy Noether Research Group (grant number NE2441/1-1).

\section*{Data Availability}
All the IllustrisTNG simulations, including TNG-Cluster are publicly available and accessible at \url{www.tng-project.org/data}, as described in \cite{nelson_illustristng_2019}.



\bibliographystyle{mnras}
\bibliography{references} 




\appendix

\section{Feature importance for convolved spectra} \label{sec:appendix_feature_importance_chandra}
We repeat the procedure of Section \ref{sec:feature_importance}, now applied to the spectra after convolution with the \textit{Chandra} response function. The network is fully retrained for these convolved spectra, so any similarities with the full resolution feature importances are independent findings. 

\begin{figure*}
    \centering
    \includegraphics[width=\linewidth]{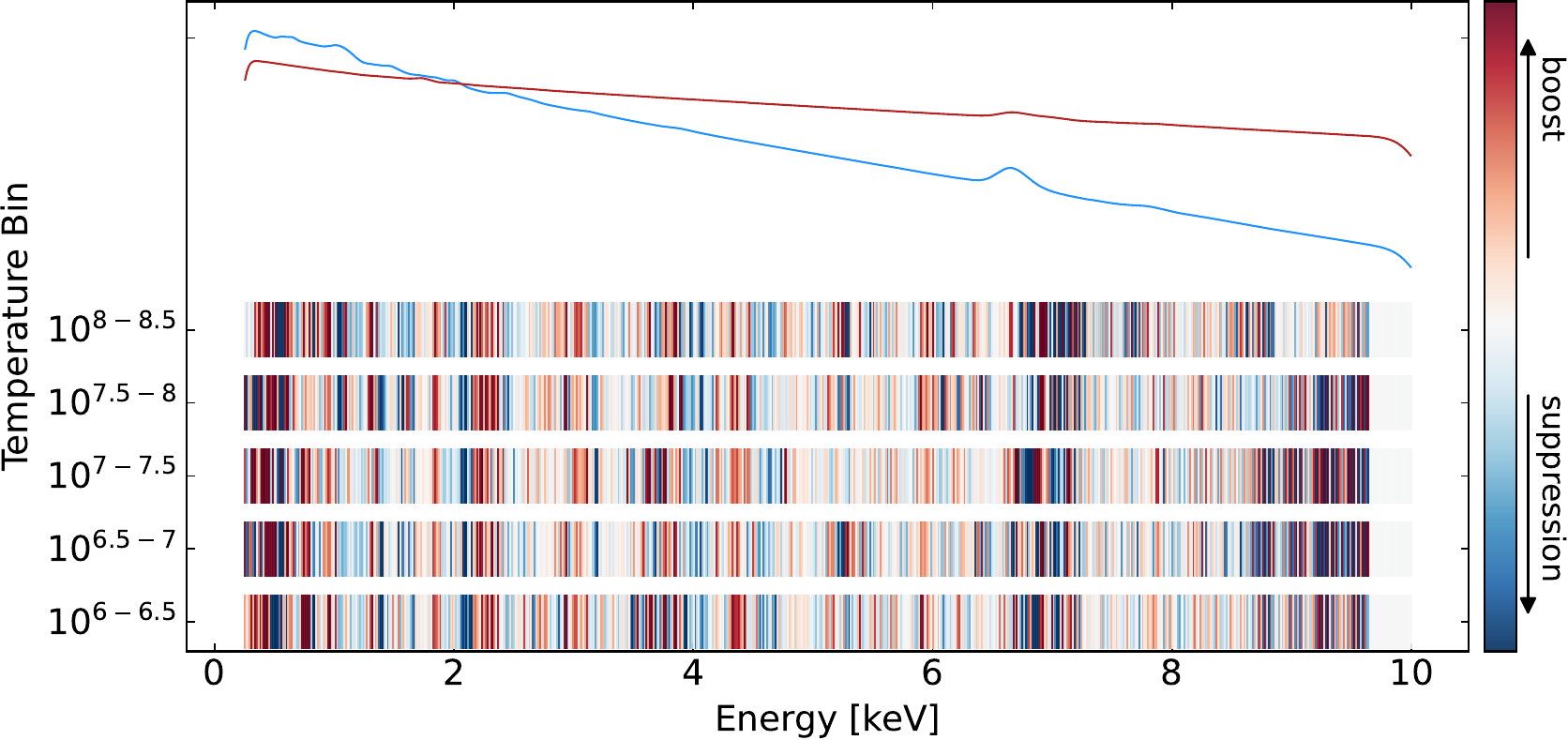}
    \caption{As Figure \ref{fig:shap_values}, but applied to the model trained on \textit{Chandra} convolved spectra.}
    \label{fig:shap_values_chandra}
\end{figure*}

The result is shown in Figure \ref{fig:shap_values_chandra}. The feature importance is less concentrated on the line rich regions, most likely because these are smoothed away. Instead, the entire spectral range is used. Of particular note is the large importance placed on the spectrum above $8 ~\mathrm{keV}$, a region that had near zero importance in the high-resolution spectra.

Computing the relative importance of the narrow- and broad-feature pathways, we surprisingly find that the narrow feature pathway dominates
\begin{align*}
    S_{\text{narrow}} = 0.88 \, , \\
    S_{\text{broad}} = 0.09 \, , \\
    S_{I} = 0.03 \, .
\end{align*}
A potential explanation is that the stride-heavy convolutional block responsible for extracting narrow features is sensitive to a window of spectral bins, rather than spectral energy. As the spectral bins in the case of a convolution with the \textit{Chandra} response are much wider, this layer is effectively already picking up on much broader features than in the high spectral resolution case. The same happens to the convolutional block focussing on broader features, which perhaps now stretches so wide that any sensible information is lost.

\bsp	
\label{lastpage}
\end{document}